\documentclass[amsmath,twocolumn,superscriptaddress,prb,aps]{revtex4-1}
\usepackage{amsthm,amsfonts,graphicx,verbatim, color}
\usepackage{graphicx}
\usepackage{bm}
\usepackage[utf8]{inputenc}
\let\oldmarginpar\marginpar
\renewcommand\marginpar[1]{\-\oldmarginpar[\raggedleft\footnotesize #1]%
{\raggedright\footnotesize #1}}

\newcommand{\be}{\begin{equation}}
\newcommand{\ee}{\end{equation}}
\newcommand{\bea}{\begin{eqnarray}}
\newcommand{\eea}{\end{eqnarray}}

\renewcommand{\epsilon}{\varepsilon}

\def\beq{\begin{equation}}
\def\eeq{\end{equation}}
\def\bea{\begin{eqnarray}}
\def\eea{\end{eqnarray}}

\begin{document}

\title{A mean-field theory of nearly many-body localized metals}
 \author{Sarang Gopalakrishnan}
 \affiliation{Department of Physics, Harvard University, Cambridge, Massachusetts 02138, USA} 
\author{Rahul Nandkishore}
\affiliation{Princeton Center for Theoretical Science, Princeton University, Princeton, New Jersey 08544, USA}
\begin{abstract}
We develop a mean-field theory of the metallic phase near the many-body localization (MBL) transition, using the observation that a system near the MBL transition should become an increasingly slow heat bath for its constituent parts. 
As a first step, we consider the properties of a many-body localized system coupled to a generic ergodic bath whose characteristic dynamical timescales are much slower than those of the system. As we discuss, a wide range of experimentally relevant systems fall into this class; we argue that relaxation in these systems is dominated by collective many-particle rearrangements, and compute the associated timescales and spectral broadening. 
We then use the observation that the self-consistent environment of any region in a nearly localized metal can itself be modeled as a slowly fluctuating bath to outline a self-consistent mean-field description of the nearly localized metal and the localization transition. In the nearly localized regime, the spectra of local operators are highly inhomogeneous and the typical local spectral linewidth  is narrow. The local spectral linewidth is proportional to the DC conductivity, which is small in the nearly localized regime. This typical linewidth and the DC conductivity go to zero as the localized phase is approached, with a scaling that we calculate, and which appears to be in good agreement with recent experimental results (Ref.~\onlinecite{Shahar}).
\end{abstract}
\maketitle

\section{Introduction}

Many-body localized (MBL) states are states of isolated, macroscopic quantum-mechanical systems in which thermal equilibration and transport are absent. MBL states differ from the more familiar ``ergodic'' states in which a macroscopic system acts as its own bath, thus bringing itself into thermal equilibrium. Since MBL states were first predicted to exist in certain disordered interacting systems~\cite{anderson_absence_1958, fleishman, gornyi_interacting_2005, basko_metalinsulator_2006, oganesyan_localization_2007, Znid, pal_many-body_2010, imbrie, arcmp}, they have been shown to exhibit a number of unusual properties, including orders and phase transitions that are forbidden in equilibrium~\cite{huse_localization_2013, pekker_hilbert-glass_2013, bahri_localization_2013, qhmbl}. The properties of MBL states, such as the absence of intrinsic decoherence and the persistence of topological order in highly excited states, make them potentially valuable resources for quantum computation. 

Much of the theoretical work on MBL to date has focused on the deeply localized regime, in which strong-disorder methods can be applied~\cite{pekker_hilbert-glass_2013, vosk_many-body_2013} and a simple phenomenological description exists~\cite{serbyn_universal_2013, *serbyn_local_2013, huse_phenomenology_2013, bauer_area_2013, huse_phenomenology_2014, kjall}. Such methods do not readily extend to the complementary regime, in which MBL \emph{emerges} from a thermalizing state; indeed, the nature of the MBL-to-ergodic transition in a closed system is at present an unsolved problem, despite recent theoretical progress on one-dimensional systems~\cite{barlev1, *barlev2, kartiek, laflorencie}.

Our aim in this work is to construct a mean-field theory of the MBL transition, using the following idea: the MBL phase differs from the ergodic phase by failing to act as a bath for its constituent degrees of freedom; correspondingly, local spectral lines are infinitely sharp in the MBL phase (owing to the absence of relaxation) and broadened in the ergodic phase. In the spirit of dynamical mean-field theory~\cite{kotliar_RMP, dobr, *dobr2}, one can separate out some subset of the degrees of freedom (namely, a single localization volume), solve for the properties (specifically, the widths of spectral lines) in the presence of a ``bath'' due to the rest of the system, and use these results to self-consistently determine the properties of the bath. 

There are naturally two parts to this problem: first, understanding the properties of a localized system coupled to a bath; and second, implementing self-consistency. We partially addressed the first question in Ref.~\onlinecite{ngh}; however, the baths considered there were ``broad-band'' and had broad featureless spectra (or, equivalently, short correlation times). By contrast, one expects that near the MBL transition (assuming it is continuous) relaxation will be extremely slow. Thus the appropriate baths in the present context are slowly fluctuating baths with \emph{long} correlation times. Thus, the first half of this paper addresses, quite generally, the behavior of MBL systems coupled to slowly fluctuating baths, and the second half builds a mean-field theory on these results. The problem of an MBL system coupled to a slowly fluctuating bath is of considerable independent interest: for instance, the dominant source of decoherence in low-temperature solid-state systems is the nuclear spin bath, which fluctuates slowly compared with the electronic degrees of freedom. More generally, slowly fluctuating baths occur intrinsically in systems that are in some sense near an MBL transition: e.g., in disordered dipolar systems~\cite{levitov1990, burin98, yao13} or in metals close to a localization transition, energy is transported through a percolating network of widely spaced resonances; the energy scales of this network are much weaker than the nearest-neighbor coupling.

We find that relaxation in the presence of a slowly fluctuating bath is qualitatively different from that in the wide-bandwidth regime~\cite{ngh}: a typical local  transition in the system requires much more energy than the bath can supply or absorb; consequently, relaxation takes place through collective rearrangements, which involve increasingly many particles as the bath bandwidth decreases.
The associated relaxation timescale goes as a power law of the bandwidth of the bath. 
We then extend these results to construct a self-consistent mean-field theory, as outlined above. Our mean-field theory---which is valid in the regime where the thermalization time is longer than any other timescale in the system---goes beyond the original perturbative analysis of Basko et al.~\cite{basko_metalinsulator_2006} by including processes by which local charge rearrangements cause nearby levels to jitter, through Hartree shifts. 
In this regime, the local spectrum should be highly inhomogeneous, consisting of narrowly broadened spectral lines. Each of these narrow spectral lines may be viewed as a slowly fluctuating internal bath, and thus we can use the results derived in this paper to 
construct a self-consistent mean-field description of a \emph{nearly localized metal}, viz. a thermalizing system close to the localization transition, which is characterized by a highly inhomogeneous local spectrum.

The nearly localized metal is characterized by local operators having a spectrum consisting of lines with a broadening $\Gamma$. The relaxation timescale is of order $1/\Gamma$. We determine how $\Gamma$ scales with various parameters (particularly the temperature), and calculate how $\Gamma$ scales to zero as the localized phase is approached. Moreover, $\Gamma$ can be related to quantities such as the DC conductivity. We predict that, in the nearly-localized regime, the low-temperature DC conductivity vanishes with a faster-than-activated temperature dependence. Over an intermediate temperature regime, the conductivity should scale as 
\begin{equation}
\sigma \sim \exp(-a' T_0/(T-T_0)) \label{scaling}
\end{equation} 
consistent with the predictions of Ref.\onlinecite{agkl}. However, close enough to $T_0$, the conductivity should scale to zero much more slowly, i.e. close to the critical point, the metal should be parametrically more stable than predicted by Ref.\onlinecite{agkl}. These predictions appear consistent with recent experimental results (Ref.\onlinecite{Shahar}).

While the near critical regime is analytically inaccessible, a numerical solution of the mean field equations can be obtained. A graphical fit to the numerical solution suggests that the linewidth in the near-critical regime may scale as 
\begin{equation}
\sigma \sim \exp(- a'' / (T-T'_0)^{1/3})
\end{equation}
which is indeed slower than Eq.\ref{scaling}. While it is unclear how general this scaling is, our analysis suggests that it should be universally true that the near-critical scaling is much slower than the single-quantum dot scaling (\ref{scaling}) that holds further from the critical point.

Our discussion is arranged as follows. In section II, we discuss the relaxation of an MBL system coupled to a slowly fluctuating ``external'' bath. 
In section III we enumerate various experimental instances of MBL systems coupled to slowly fluctuating external baths. In section IV, we turn to the case where the MBL system is coupled to a \emph{self-consistent} (i.e., ``internal'') bath with slow dynamics: we first introduce a solvable zero-dimensional model~\cite{agkl} (see also Ref.~\onlinecite{laumann2014}) that contains the physics of many-particle rearrangements, and then use this model as the basis for a mean-field description of a nearly localized metal~\footnote{We note that there are some similarities between Section IV and Ref.\onlinecite{Fiegelman}. However, \onlinecite{Fiegelman} is concerned with the localization transition for single excitations above the ground state, whereas we are concerned with the very different problem of the localization transition for generic high energy states. Moreover, unlike \onlinecite{Fiegelman} we work on a regular lattice (not a Bethe lattice), and incorporate Hartree terms; as we discuss, this changes the nature of metallic-state solutions.}. We introduce a self consistency equation the solutions to which capture both the localized phase and the metallic phase near localization. In section V, we discuss the solutions to this self consistency equation, and hence deduce how the line width scales to zero as the localized phase is approached.

\section{Slowly fluctuating external bath}\label{external}

\subsection{Model}

\begin{figure}[tb]
\begin{center}
\includegraphics{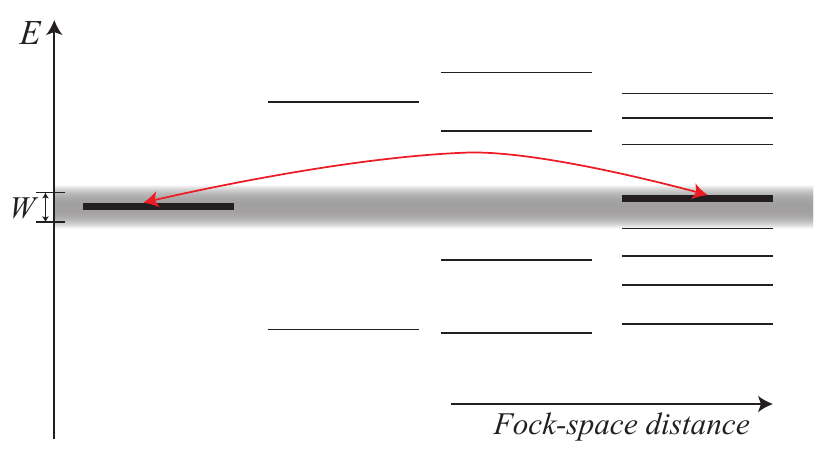}
\caption{Schematic illustration of the physics of bandwidth-limited relaxation. Only levels within $W$ of each other can undergo bath-mediated transitions; however, levels that are nearby in energy are typically far apart in real space and/or Fock space.}
\label{fig1}
\end{center}
\end{figure}

We begin by discussing bandwidth-limited relaxation in systems where all many-body eigenstates are localized (the ``fully many-body localized'' or FMBL regime); such systems have a simple phenomenological description (the ``l-bit'' model \cite{huse_phenomenology_2013, bauer_area_2013, *serbyn_local_2013, huse_phenomenology_2014}). We assume that our system+bath Hamiltonian is given by $H = H_0 + H_{bath} + H_{int}$, where $H_0$ is the l-bit Hamiltonian for the MBL system\cite{huse_phenomenology_2014}, 
\begin{equation}
H_0 = \sum_i h_i \sigma^z_i + \sum_{ij} J_{ij} \sigma^z_i \sigma^z_j + \sum_{ijk} V_{ijk} \sigma^z_i \sigma^z_j \sigma^z_k  + \ldots
\end{equation}
Here, $\sigma^z_{ij}$ are spin-1/2 degrees of freedom (l-bits) that live on a $d$-dimensional lattice. The various $z-z$ couplings are short ranged (potentially with exponential tails), and the many body eigenstates are eigenstates of all the $\sigma^z_i$. The $\sigma^z_i$ are the emergent local conserved quantities, which are related to the physical degrees of freedom by a local unitary transformation of finite depth (equal to the localization length $\xi$), with potential exponential tails. Flipping a single l-bit typically changes the energy by an amount $\Delta$, the characteristic energy scale of the system. 

We now couple the l-bit Hamiltonian to a generic, thermalizing bosonic bath of bandwidth $W$; the bosons are taken to live on the links of the original lattice. The system-bath coupling takes the $\sigma^z$-conserving form 

\beq\label{lbit-int}
H_{int} =\gamma \sum_{\langle i j \rangle} (b^\dagger_{\langle i j \rangle} + b_{\langle i j \rangle}) (\sigma^+_i \sigma^-_j + h.c. + \ldots)
\eeq
where the ellipses denote long-range and/or high-order jumps that fall off exponentially with distance, with decay length $\xi$. 

Before turning to the narrow-bandwidth limit, we briefly summarize the results~\cite{ngh} for the wide-bandwidth limit  $W \gg \Delta$. In this limit, the factors limiting relaxation are the (weak) system-bath coupling $\gamma$ (referred to in Ref.\onlinecite{ngh} as $g$), and the temperature of the bath, $T$. At low temperatures, $T \ll \Delta$, most spins are frozen in their ground-state configuration; excitations are essentially single-particle in character, and relaxation takes place as in a non-interacting localized state. At high temperatures $T \gg \Delta$, again, the relaxation dynamics shows few signatures of many-body processes: the bath is able to place nearest-neighbor l-bit hops on-shell, and these processes therefore dominate relaxation.

These considerations suggest that nontrivial relaxation is likeliest to occur in the parameter regime $W \ll \Delta \ll T$, such that the system and the bath are effectively at high temperature, but the bath is narrow bandwidth. In the rest of the present section, we shall focus on this regime; we shall further assume that $\gamma \ll W$, so that the bath behaves in an effectively Markovian fashion on the timescales over which the system couples to the bath. Finally, we assume that the bath has a higher heat capacity than the system; for narrow-band baths this entails the assumption that 
\beq\label{schottky}
\mathcal{N} (W^2/T^2) \gg 1
\eeq
where the bath has $\mathcal{N} \gg 1$ times as many degrees of freedom as the system. 

\subsection{Relaxation channels}\label{channels}

We now discuss three kinds of processes by which a slowly fluctuating bath can induce relaxation in the l-bit model. The general principle is as follows: because $W \ll \Delta/z$ (where $z$ is the coordination number), the bath cannot supply or absorb the energy for a single-l-bit hop. Long-range or high-order processes connect the initial state to many more final states (i.e., have a smaller accessible level spacing), but have a smaller matrix element. For simplicity we shall work with a bounded bath spectrum having uniform spectral density on the energy interval $[-W/2,W/2]$. The results generalize straightforwardly to cases in which the bath has a Gaussian or other sufficiently sharply peaked spectrum; for a Lorentzian spectrum, however, the dominant processes are typically nearest-neighbor hops that exploit the tails of the Lorentzian (see Appendix). For the bounded bath spectrum, it is clear that the dominant relaxation processes are those for which the effective level spacing is $\sim W$; in what follows, we shall estimate the rates of these processes [Eqs.~\eqref{singleLbit}, \eqref{manyLbit}, \eqref{highorder}]. We find that, as $W \rightarrow 0$, relaxation is dominated by multiple-l-bit rearrangements~\eqref{manyLbit}. The mechanisms described in this section are analogous to variable-range hopping~\cite{mottVRH}, in either real space or Fock space; because we focus on high temperatures and short-range interactions, the physics of the Coulomb gap is not relevant to our analysis. We note also that our argument is at the `Mott variable-range hopping' level of sophistication - more rigorous results could in principle be obtained via percolation theory, but such calculations are beyond the scope of the present analysis. 

\subsubsection{Long-range single l-bit hop}
 The coupling~\eqref{lbit-int} includes exponentially suppressed long-range hops of the form $\gamma (b + b^\dagger) \exp(-|\mathbf{r}_i - \mathbf{r}_j|/\xi) \sigma^+_i \sigma^-_j$. To find a single l-bit hop that is on shell to precision $W$, we must typically hop a distance $R$, where $\Delta R^{-d} \sim W$. This yields $R \sim (\Delta/W)^{1/d}$, and the matrix element for a hop at this range is $\gamma \exp\big(-(\Delta/W)^{1/d}/\xi\big)$. Thus the relaxation rate is

\beq\label{singleLbit}
\Gamma_a \sim \frac{\gamma^2}{W} \exp\big(-2(\Delta/W)^{1/d}/\xi\big).
\eeq

\subsubsection{Multiple-l-bit rearrangements}
In addition to long-range hops, Eq.~\eqref{lbit-int} includes terms of the form $\gamma (b + b^\dagger) \sum \sigma^+_i \sigma^-_j \sigma^+_k \ldots$ (which are analogous to the many-body rearrangements discussed later in the zero-dimensional case). In order to place the system on shell to within $W$ one must generically rearrange all the l-bits in a volume of linear dimension $\hat{n}$, where $\Delta \exp \big(-s(T) \hat{n}^d\big) \sim W$. Here $s(T)$ is the entropy per l-bit, which interpolates between the limits $s(\infty) = 1$ and $s(0) = 0$, and measures the fraction of spins that are free to flip (see discussion in Ref.\onlinecite{ngh} for details). Because excitations are localized in Fock space as well as real space (see Sec. IV below, as well as Refs.~\onlinecite{agkl, basko_metalinsulator_2006}), the corresponding matrix element is  $\sim \gamma \exp(-\hat{n}/\xi - s(T) \hat{n}^d/\Xi)$, where $\xi, \Xi$ are the real-space and Fock-space localization lengths respectively. In $d > 1$, for sufficiently small $W$, this is to leading order

\beq
\Gamma_b \sim \frac{\gamma^2}{W} \left(\frac{W}{\Delta} \right)^{1/\Xi}
\label{manyLbit}
\eeq
This power-law dependence holds more generally in $d = 1$, but with $\Xi$ replaced by $\xi \Xi / (\xi + \Xi)$.

\subsubsection{Higher-order coupling to bath}
 A third channel for relaxation involves going to high order in the system-bath coupling, rearranging all l-bits within a volume of linear dimension $\hat{n}\ge n_c$, where $\Delta \exp \big(-s(T) n_c^d\big) \sim n_c^d W$. The approximate solution is when $\hat{n} \sim \left(\frac{1}{s(T)}\ln(\Delta/W)\right)^{1/d}$. This gives a relaxation rate (for large $\hat{n}$)

\beq
\Gamma_c  \sim \frac{\gamma^{2\hat{n}^d}}{W \Delta^{2\hat{n}^d-2}} \sim \frac{\gamma^2}{W} \left(\frac{W}{\Delta}\right)^{\frac{2\ln(\Delta/\gamma)}{s(T)}}
\label{highorder}
\eeq
where in the first line we have assumed that all the intermediate virtual states are off-shell by the typical amount $\Delta$. 

In the limit $\gamma/\Delta, W/\Delta \rightarrow 0$, the leading relaxation rate is $\Gamma_b$, given by Eq.(\ref{manyLbit}).

\subsection{Spectral Lineshape}\label{external:spectral}

We now use the estimate~\eqref{manyLbit} of the relaxation rate in the $\gamma, W \rightarrow 0$ limit to discuss the l-bit spectral lineshape. The definition of a ``lineshape'' in this context is ambiguous: for instance, even for vanishing system-bath coupling, the frequency of each l-bit line depends on the state of all the other l-bits, so upon thermal averaging the l-bit frequency can in principle be continuous and independent of the system-bath coupling~\cite{ngh}. This ``inhomogeneous'' broadening is not of interest to us; it can be reversed by spin echo~\cite{mblDeer}, and does not correspond to real ``irreversible'' dynamics. Rather, we are interested in the linewidth corresponding to the spin echo decay rate~\cite{mblDeer}; this linewidth indeed vanishes as $\gamma \rightarrow 0$. 

\begin{figure}[tb]
\begin{center}
\includegraphics[width=0.45\textwidth]{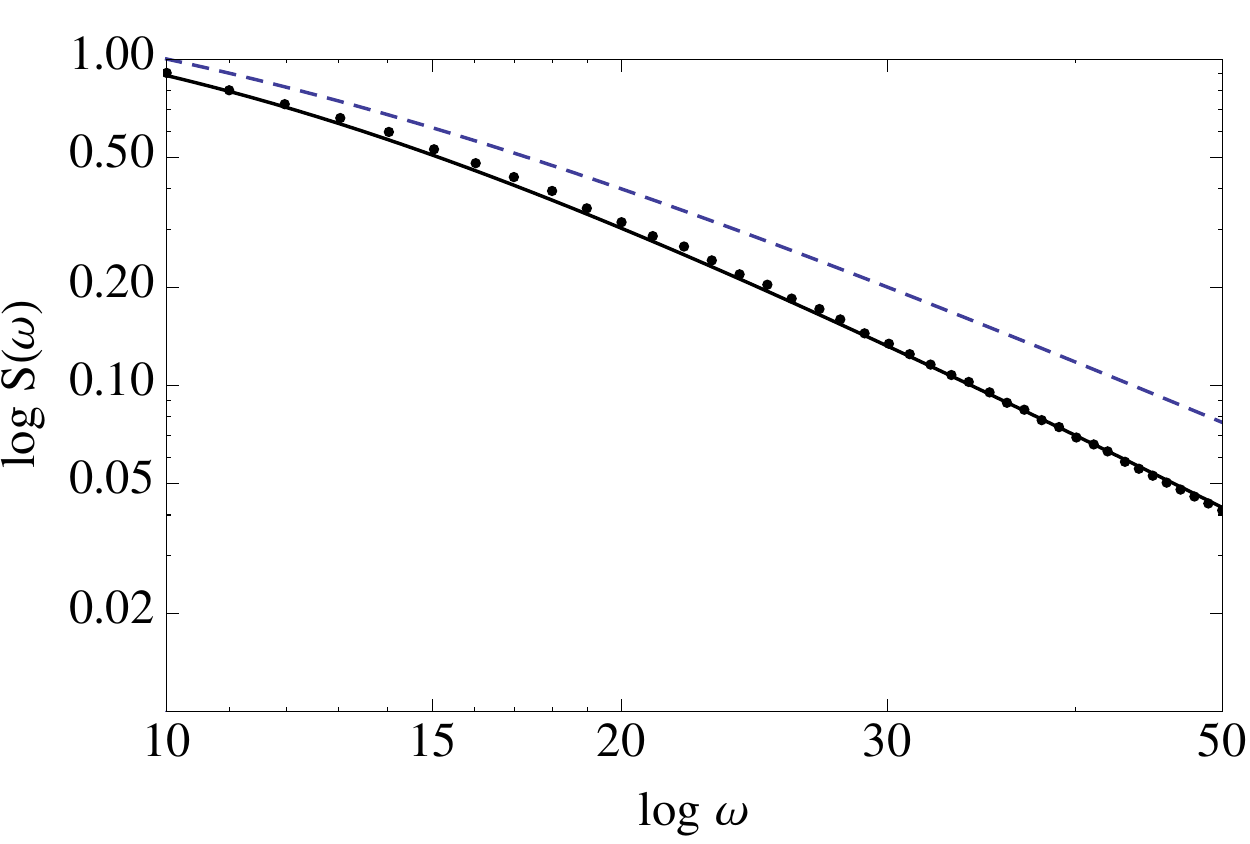}
\caption{Numerical Fourier transform of the decay~\eqref{echo} (dots) and analytic approximation~\eqref{approxline} (solid) on a log-log plot, for parameters $\Gamma_b = 1, \xi^d s(T) = 4, J = 1000$. Approximating the lineshape with a Lorentzian (dashed line) would lead to an overestimate of the weight in the spectral tails.}
\label{lineshapefit}
\end{center}
\end{figure}

There are two contributory processes to this linewidth. First, each l-bit flips at a rate $\Gamma_b$; as the density of final l-bit states is constant on energy scales $\alt \Delta$, a Golden-Rule argument shows that this decay is exponential on timescales $t > 1/\Delta$. Second, the flipping of each l-bit causes the energies of nearby l-bits to fluctuate. The echo decay of a spin coupled to fluctuating two-level systems was exactly solved in Ref.~\onlinecite{galperin}; we quote the asymptotic forms for the spin-echo response $\psi_i(t)$ at a site $i$

\beq
\psi_i \sim \prod_{j} \psi_{ij}; \quad \psi_{ij}(t) = \left\{  \begin{array}{cc} 1, & \hat{J}_{ij} t \ll 1 \\ e^{-\Gamma_b t}, & \hat{J}_{ij} t \gg 1 \end{array}  \right.
\eeq
where the product is over all sites $j$ such that the effective interaction $\hat{J}_{ij} \agt \Gamma_b$. Further-away, more weakly coupled spins contribute negligibly. Combining these two processes, we find that the spin echo decay rate is

\beq\label{echo}
\psi(t) = \exp(-\Gamma_b n_{fast}(t) t)
\eeq
where

\beq\label{nfast}
n_{fast}(t) = \left\{\begin{array}{cc} 1 & t \ll 1/J \\ s(T) \xi^d \ln^d(J t) & 1/J \ll t \ll 1/\Gamma_b \\ s(T) \xi^d  \ln^d(J/\Gamma_b) & t \gg 1/ \Gamma_b\end{array} \right.
\eeq
where $J$ is the nearest neighbor interaction and further neighbor interactions fall off exponentially with decay length $\xi$ \cite{huse_phenomenology_2014}. Fourier-transforming this expression gives the effective lineshape. To a good approximation (see Fig.~\ref{lineshapefit}), the lineshape is given by

\beq\label{approxline}
S(\omega) \simeq \frac{\Gamma_b n_{fast}(1/\omega)}{\omega^2 + (\Gamma_b n_{fast}(1/\omega))^2}
\eeq

Note that, when $\Gamma_b \ll J$, this deviation from Lorentzian behavior becomes more pronounced. The width of the central peak is $\tilde \Gamma$, defined by
\begin{equation}
\tilde \Gamma \approx \Gamma_b n_{fast}(1/\tilde \Gamma) \label{gamma}
\end{equation}
To leading order in small $\Gamma_b$, this yields $\tilde \Gamma \sim \Gamma_b s(T) \xi^d \ln^d(J/\tilde \Gamma_b)$, consistent with Ref.~\onlinecite{ngh}.

We note in passing that an additional spectral feature is possible if the system-bath coupling has diagonal (i.e., pure dephasing) terms of the form $\sigma^z (b^\dagger + b)$, and if the detuning $\delta \omega < W$. In this special case, the system can put itself `on shell' by borrowing energy from the bath without needing to undergo a collective rearrangement. For concreteness let us assume a tip-to-sample matrix element $t$, which is much smaller than any other scale in the problem. Absorption of a quasiparticle then proceeds via an intermediate virtual state with detuning $\delta \omega$, where the matrix element to go to the virtual state is $t$ and the matrix element to leave the virtual state by borrowing energy from the bath is $\gamma$. Given that the spectral density in the bath is $1/W$, it follows that tunneling proceeds at a rate $t^2\gamma^2/\delta \omega^2 W$, such that the tunneling density of states is $\gamma^2/\delta \omega^2 W$, in the special rase that $\delta \omega < W$, and diagonal couplings to the bath are allowed.

To summarize, although the dependence of the relaxation rate and linewidth on $W$ are very different in the wide- and narrow-bandwidth limits, the phenomenology {\it as a function of $\tilde \Gamma$} is identical to the discussion in Ref.\onlinecite{ngh}, and results obtained in Ref.\onlinecite{ngh} can be carried over {\it mutatis mutandis} to the narrow-band bath case, changing only $\Gamma(g)$ to $\tilde \Gamma(W)$ given by Eq.~(\ref{gamma}). We recall in particular that the DC conductivity $\sigma$ is related to the local linewidth $\tilde \Gamma$ by $\sigma \sim \tilde \Gamma$.

\subsection{Intermediate system-bath coupling}\label{intcoupling}

We now comment briefly on how these arguments would be modified in the regime of intermediate system-bath coupling, $W \ll \gamma \ll \Delta$, as is typically the case for hyperfine interactions~\cite{khaetskii1, khaetskii2, witzel1, *witzel2, bloembergen, *szabo, paola2009}. In this regime our Golden Rule arguments cannot be used directly because of the importance of the back-action of the system on the bath. Let us consider, for concreteness, a single (``central'') system spin at the origin of coordinates, coupled to a lattice of $\mathcal{N}$ bath (e.g., nuclear) spins, with a coupling of the form $\gamma \mathbf{S} \mathbf{\cdot S}_{bath}(j) \exp(-j)$. The bath spins are taken to have local interactions such that the bandwidth is $W$. One can see immediately that the bath spins within $\ln(\gamma/W)$ of the origin get locked to
 the central spin; these nearby spins are called the ``frozen core'' in nuclear magnetic resonance~\cite{bloembergen, *szabo, paola2009}, and should be included in the system rather than the bath. For bath spins that are far away, on the other hand, the effective system-bath coupling is $\gamma_{eff} \alt W \ll \gamma$, allowing us to apply our previous analysis.

\section{Experimental relevance}
The most common instance of an ergodic, slowly fluctuating external bath in solid-state systems is a nuclear spin bath. The dynamics of such a bath are due to nuclear spin diffusion, $H_{bath} \sim \sum\nolimits_{ij} J_{ij} \mathbf{I}_i \mathbf{\cdot I}_j/|\mathbf{r}_i - \mathbf{r}_j|^3$, where the bandwidth associated with $J_{ij}$ is on the order of $10$ kHz~\cite{khaetskii1, khaetskii2, witzel1, *witzel2}, which is orders of magnitude smaller than the typical electronic scales. Under typical experimental conditions~\cite{witzel1, *witzel2}, the nuclear spin bath is effectively at infinite temperature. Thus, realizations of MBL that use the electronic spins of (e.g.) nitrogen-vacancy centers in diamond~\cite{yao13} are naturally subject to narrow-band relaxation due to this nuclear spin-bath at low temperatures. (Indeed, power-law dephasing of the electron spins was predicted on unrelated grounds in Refs.~\cite{khaetskii1, khaetskii2}.) We should caution, however, that making quantitative predictions for such systems would require one to address various issues specific to the nuclear case that are beyond the scope of this paper.
A more \emph{tunable} kind of narrow-bandwidth bath can be realized by placing the MBL system in a high-$Q$ resonator; in this case, the bandwidth of the bath is set by the linewidth of the resonator mode. Our results are also applicable to the experimentally relevant situation of almost MBL states in systems with long-range interactions, $V(R) \sim 1/R^\alpha$ (e.g., dipoles in three or possibly two dimensions~\cite{yao13}). In these settings, most of the degrees of freedom are almost localized; however, there is a thermalizing network of long-range resonances at a scale $R_0$, which can be regarded as an internal bath whose bandwidth, $\sim 1/R_0^\alpha$, is narrow in the strong-disorder limit. Finally, our results can be applied to almost localized systems coupled to phonons or other Goldstone modes, if all phonons at frequencies $\omega > \omega_c$ are localized, with $\omega_c$ suitably small (as in Ref.~\onlinecite{ariel_phonon}).

\section{Slowly fluctuating ``internal'' bath: the nearly localized metal}
 The central idea in this section is that while the local spectral function deep in the metallic phase is smooth, the local spectral function in the localized phase is made out of delta functions and contains a hierarchy of gaps \cite{ngh}. Thus as the localization transition is approached the local spectral function in the metal should become increasingly inhomogeneous, and should be composed of increasingly narrow lines. We apply the ideas of the previous section to compute the width of a typical spectral line coupled to a self-consistent bath in a typical-mean-field approximation (similar in spirit to Refs.~\cite{kotliar_RMP, dobr, *dobr2}), and hence obtain a mean-field description of the nearly localized metal.
 
\subsection{Approach}

In previous sections we discussed the behavior of a system coupled to an external bath with slow dynamics. However, any region of an isolated quantum system in the thermalizing regime also behaves as if it were connected to an `internal' heat bath, and close to the localization transition this `internal' heat bath should have asymptotically slow dynamics. 
In what follows we use this observation to develop a simple mean-field theory of the metallic phase near a many-body localization transition. This mean-field theory allows us to compute the MBL-to-ergodic phase boundary, and to estimate how the linewidth vanishes as the MBL transition is approached.


We work with a model that is closely related to that of Ref.~\onlinecite{basko_metalinsulator_2006}. This model is specified in terms of electrons hopping on a lattice with short range interactions, such that electrons within a single-particle correlation length $L$ (in real space) are strongly coupled, but electrons further apart than $L$ (in real space) are weakly coupled. It is important that $L$ be much larger than the lattice scale. The system is now carved up into correlation volumes, each of size $L$, and each correlation volume is treated using a zero-dimensional model of an interacting quantum dot, which we treat in the manner of Ref.~\onlinecite{agkl}. The analysis proceeds as follows: in Sec.~\ref{qdot} we compute the relaxation rate of each quantum dot, subject to a slowly fluctuating external bath; in Sec.~\ref{baasetup} we specify the nature of the interdot coupling; and in Sec.~\ref{selfconsist} we replace the external bath from Sec. \ref{qdot} by a self consistent internal bath, and calculate the relaxation rate due to the interdot coupling. This gives us a self-consistent mean-field equation for the linewidth, to which we then present solutions in the next section. 
While we shall use the terminology of (implicitly quantum-mechanical) ``baths'' throughout, our calculations can be reinterpreted in terms of a quantum dot coupled to a self-consistent classical noise field~\cite{ariel-diffusion},\footnote{A. Amir, in preparation}. 

\subsection{Quantum dot coupled to slowly fluctuating bath}\label{qdot}

The first step in constructing our mean-field theory is to estimate the relaxation rate of a level in the zero-dimensional model of Ref.~\onlinecite{agkl} in the presence of a slowly fluctuating (external) bath. We therefore begin by reviewing some results on this model, which consists of a quantum dot with dimensionless conductance $g = E_T/\Delta$, where $E_T$ is the Thouless energy of the dot (i.e., the rate at which electrons diffuse across the dot) and $\Delta$ is the single-particle level spacing~\footnote{In a homogeneous system, the MBL transition occurs at $g \alt 1$; however, as we are considering a system of coupled grains, we may consistently assume that $g \gg 1$.}. The electronic states on the dot are described by the Hamiltonian~\cite{agkl, aleiner-review}

\beq\label{H-agkl}
H_{\mathrm{AGKL}} = \sum_{\alpha = 1}^N \varepsilon_\alpha c^\dagger_\alpha c_\alpha + V_{\alpha\beta\gamma\delta} c^\dagger_\alpha c^\dagger_\beta c_\gamma c_\delta
\eeq
where the statistics of $\varepsilon_\alpha$ and of $V_{\alpha\beta\gamma\delta}$ are specified, up to an overall interaction strength $\lambda$, by random-matrix theory~\footnote{Note that, although the \emph{single-particle} properties are described by random-matrix theory, in general the \emph{many-body} level statistics are not.}. 

We now briefly review the closed-system behavior of the Hamiltonian~\eqref{H-agkl}, focusing on its eigenstates. For $\lambda = 0$, each eigenstate is  parameterized by the occupation numbers of $N$ single-particle levels; thus, it forms a vertex of an $N$-dimensional hypercube (with quenched on-site disorder). The interaction term in $H_{\mathrm{AGKL}}$ changes only four occupation numbers at once, and therefore acts as a local hopping term on the hypercube. Ref.~\onlinecite{agkl} argued, by mapping the Fock hypercube onto a Cayley tree with coordination number $K \equiv g^3/6$, that the many-body eigenstates undergo a transition as a function of $\lambda$ between a low-energy localized regime (in which the eigenstates are localized in Fock space, in the sense that their amplitudes at a point on the Fock hypercube decrease exponentially with distance from some central site~\cite{abouchacra, *chalker}) and a high-energy delocalized regime (sufficiently deep in this delocalized regime, the eigenstates are ``ergodic,'' i.e., spread evenly over all configurations with the appropriate energy). The transition point can be approximately estimated by comparing the typical matrix element $\lambda$ to the accessible level spacing at a given energy $\delta_m(\varepsilon) \simeq \Delta^3/\varepsilon^2$. For sufficiently small $\lambda$, all eigenstates on the dot are localized. We shall assume in what follows that we are in this regime, where the isolated quantum dot is fully many body localized. The localization length in Fock space will in general depend on energy. To treat this energy-dependence correctly one would have to go beyond the Cayley-tree model~\cite{beenakker, *kamenevQD}. For simplicity, we ignore the energy dependence of the Fock space localization length, and work with a quantum dot on which all states are localized with a single characteristic Fock-space localization length $\Xi$.

We now imagine coupling the dot model~\eqref{H-agkl} to a generic, ergodic bath. The bath-mediated interaction is taken to have the particle-number-conserving form 

\beq
H_{int.} = \gamma \sum_{\alpha, \beta} c^\dagger_\alpha c_\beta (b^\dagger + b),
\eeq
where the $b$'s are the excitations of the bath, which we assume to be bosonic. The bath excitations are taken to have a bounded spectrum in the range $[E - W/2, E + W/2]$, where in general $E \neq 0$. 

We now compute the relaxation rate of an excited particle due to hopping, to leading order in small $\gamma/W$. 
 At $W=0$, a system prepared at a particular vertex of the Fock hypercube remains localized near that vertex. At non-zero $W$, the system can relax by hopping to vertices with a different value of the system energy $\epsilon$, by borrowing the missing energy from the bath. However, the bath can only supply or absorb energies that are within a range $W$ of $E$, while neighboring vertices on the Fock hypercube have a level spacing $\delta_m \gg W$, so that nearest-neighbor hops are not, in general, on shell to within $W$. To find a vertex that differs in energy by less than $W$, a particle must tunnel to the $\hat{n}$th neighbor, where $\delta_m K^{-\hat{n}} \le W$ i.e. $\hat{n} \ge n_c = \ln(\delta_m/W)/\ln K$. However, the matrix element for a long range hop is of order $\exp(-\hat{n}/\Xi)$, because of the exponential localization of states on the Fock hypercube. Summing over all distances $n \ge n_c$ in a saddle point approximation, we find that the Golden-rule estimate of the relaxation rate $\Gamma$ is 

\beq
\Gamma(W) = \frac{\gamma^2}{W} \left( \frac{W}{\delta_m} \right)^{2/(\Xi \ln K)} \label{eq: relaxation rate}
\eeq
This defines a timescale for relaxation $t_{relax} \sim \gamma^{-2} W^{1 - \frac{2}{\Xi \ln K}}$, so the relaxation time diverges as $W \rightarrow 0$. Physically, this divergence of $\hat{n}$ and the consequent slow relaxation arise because finding a transition which is on shell to precision $W$ requires increasingly large rearrangements of the system. 
Once $W$ [Eq.~(\ref{eq: relaxation rate})] becomes smaller than the total many body level spacing of the dot (which is exponentially small in $N$), then the dot reverts to being fully localized despite the coupling to the bath, since we can no longer find transitions that are on shell to a precision $W$. 

We note that the discussion above applies for  a Lorentzian bath only when $\Xi \ln K > 1$; otherwise (see Appendix) relaxation is dominated by lowest-order processes that exploit the tails of the Lorentzian. The two cases $\Xi \ln K > 1$ and $\Xi \ln K \leq 1$ are both taken into account in what follows. 

\subsection{Setup of the coupled-dot problem}\label{baasetup}

\begin{figure}[t]
\begin{center}
\includegraphics{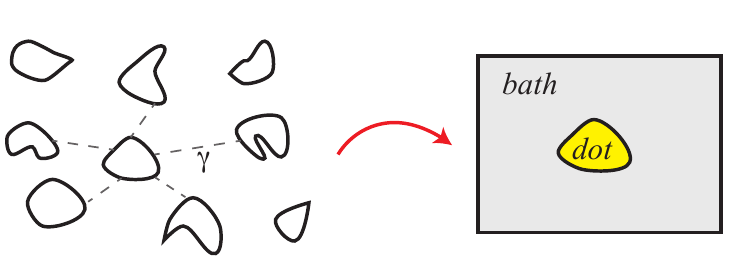}
\caption{Schematic description of our mean-field approach. We consider a lattice of statistically similar multilevel quantum dots, and approximate this as a single quantum dot coupled to a slowly fluctuating bath.}
\label{fig2}
\end{center}
\end{figure}

We now make use of the insight that the AGKL quantum dot model \cite{agkl} can be regarded as an effective description for a region with linear dimension of order the correlation length $L$, which sees a bath due to the other quantum dots to which it couples through the interaction (Fig.~\ref{fig2}) ~\footnote{The l-bit model of the FMBL state is unsuitable for our present purposes, because it has a decay length for the interactions which diverges near the localization transition, and we want a model where the interactions are short range. }. It is important for our purposes that the Fock space for each quantum dot be large, i.e., that $L$ be large, for reasons that will become apparent shortly.  A similar setup was employed in \cite{basko_metalinsulator_2006} when establishing stability of the MBL phase; however, we are not aware of any work that uses such an approach to describe the regime near the MBL transition.

\begin{figure}[b]
\begin{center}
\includegraphics{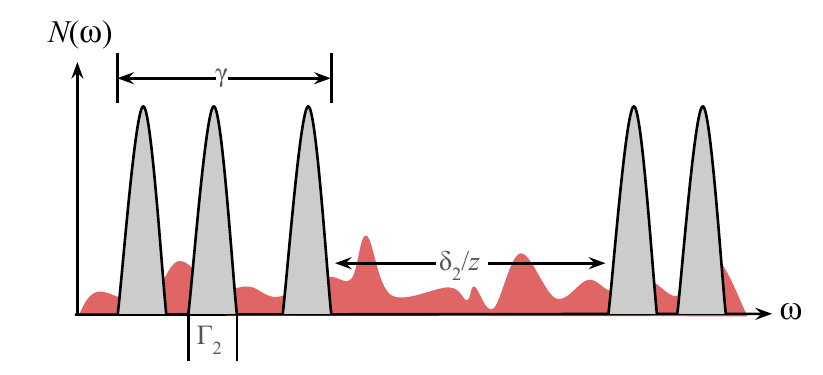}
\caption{Spectrum of the self-consistent bath, averaged over a time $1/\Gamma_1$. Due to spectral diffusion, this bath consists of clumps of $\Gamma_2/\Gamma_1$ Lorentzians, each of linewidth $\Gamma_2$, spaced within an interval $\gamma \gg \Gamma_2$ (see main text, section IVD). These clumps of Lorentzians are separated from one another by a spacing $\delta_2 \gg \gamma$. The satellite peaks contribute a broad weak background, indicated in pink.}
\label{scales}
\end{center}
\end{figure}

Following Ref.~\onlinecite{basko_metalinsulator_2006}, then, we consider an array of statistically similar quantum dots coupled through a number-conserving interaction of the form $\gamma/\sqrt{zN^2} \sum_{IJ} \sum_{\alpha \beta \gamma \delta} \mathcal{K}_{IJ} c^\dagger_\alpha(I) c_\beta(I) c^\dagger_\gamma(J) c_\delta(J)$, where $I$ and $J$ label different quantum dots, $z$ is the coordination number of the array, $N$ is (as before) the number of single-particle levels on each quantum dot, and the coupling $\mathcal{K}_{IJ}$ decays exponentially with distance between dots, with a decay length of order the size of the dots. We now assume a mean-field decoupling in which dot $I$ is self-consistently coupled to an environment consisting of all the other dots: i.e., $\gamma/\sqrt{zN^2} \sum_{IJ} \mathcal{K}_{IJ} c^\dagger_\alpha(I) c_\beta(I) c^\dagger_\gamma(J) c_\delta(J) \sim \gamma/\sqrt{zN^2} \sum_{I \alpha \beta} c^{\dag}(I)_{\alpha} c(I)_{\beta} m(I)$, where $m(I) = \sum_{J \gamma \delta} K_{IJ} c^{\dag}_\gamma(J) c_\delta(J)$. 

We now ask whether states on the dot can relax by `borrowing' energy from the environment. The environment consists of all allowed transitions in the other quantum dots. If the other quantum dots are localized, then the spectral function of the environment is made up of delta functions. In this case we will not be able to find a transition in the environment that places the transition in the dot exactly on shell. As a result, a localized quantum dot placed in a localized environment will remain localized. In contrast, if the spectral lines in the environment are slightly broadened with a linewidth $\Gamma$, then the dot will be able to relax, provided it can find a state, on shell to a precision $\Gamma$, to which it can decay. Our objective is to obtain a self consistent mean field theory that describes how this linewidth $\Gamma$ goes to zero as we approach the localized regime. 

An important quantity for our analysis is the spectral function of the self consistent environment. The maximum amplitude in the spectral function of the environment will be wherever there is a spectral line corresponding to single particle transitions on a neighboring dot. The spacing between these peaks is $\delta_2/z$, where $\delta_2 \equiv \Delta^2/ \varepsilon$ and $1/\delta_2$ is the accessible two-particle DOS on a single quantum dot. In order for our discussion to make sense, we require that these dominant spectral lines do not overlap. This then requires that $\gamma \ll \delta_2/z$, for reasons that will shortly become clear. We assume the interactions are sufficiently weak that we are operating in this regime. 

In addition to the large but sparse features coming from single particle transitions on the neighboring dots, there will be additional features coming from multi-particle transitions.  These `satellite' peaks will be much denser than the large peaks coming from single particle transitions on neighboring dots, but they will have an amplitude that is much smaller. The amplitude for peaks coming from multi particle transitions will fall off exponentially with the order of the transition (with decay constant $\Xi$). 
The spectral function of the bath will also have satellite peaks coming from distant dots; however, the amplitude of these satellite peaks falls off exponentially with distance, whereas their density grows only as a power law of distance, so for weak coupling we may ignore the satellite peaks coming from transitions on distant dots. (The Hartree shifts from distant dots will however have to be taken into account, as we will discuss).

\subsection{Self-consistency equations}\label{selfconsist}

We shall now derive self-consistent mean-field equations for the model specified in the previous section. This derivation takes place in two steps. First, we assume that a typical transition in the system happens at a rate $\Gamma$, and estimate the lineshape $S_\Gamma(\omega)$ of the typical transition as a function of $\Gamma$. (As discussed above, the typical spectral line is broadened because of the Hartree shifts due to nearby transitions---in the language of nuclear magnetic resonance, it includes both $T_1$ and $T_2$ processes.) Second, we compute the decay rate $\Gamma$ of a typical level in the presence of an environment with the characteristic lineshape $S_\Gamma(\omega)$. Together, these two relations constitute a self-consistent theory of relaxation in a nearly localized metal.

The first step is a straightforward application of the reasoning in Sec.~\ref{external:spectral}. A transition on any level causes the energies of nearby levels to fluctuate and thus broadens them. From Sec.~\ref{external:spectral} the broadened linewidth is given by

\beq
S_\Gamma(\omega) \simeq \frac{\Gamma n_{fast}(1/\omega)}{\omega^2 + (\Gamma n_{fast}(1/\omega))^2}
\eeq
where $\Gamma$ is the assumed typical rate (which we will later solve self-consistently for) and $n_{fast}(t)$ is given by Eq.~\eqref{nfast}. In particular, at short times $n_{fast} \simeq 1$ and at long times $n_{fast} \simeq s(T) \ln^d(\gamma/\Gamma)$. One can arrive at a naive mean-field theory by ignoring the Hartree shifts (i.e. by approximating $\Gamma$ by $\Gamma_b$). As we shall see below, this would reproduce the results of Refs.~\onlinecite{gornyi_interacting_2005, basko_metalinsulator_2006, Fiegelman}. A central result of our paper is that including the Hartree interactions parametrically enhances the stability of the metallic phase.

The next step is to compute the decay rate given the lineshape. We shall use the Golden Rule for this estimate. The validity of the Golden rule is discussed at length in Appendix B: briefly, our application of the Golden Rule is valid provided (a) we are sufficiently close to the localization transition that the lifetime $\Gamma$ is much smaller than the other characteristic scales, (b) we use \emph{typical} rather than thermally averaged spectral functions, and (c)~the coordination number is large enough that the ``back-action'' on an individual line can be treated in mean-field theory. Using the Golden Rule, the decay rate of a single level coupled to a self-consistent environment is given by 
\begin{equation}\label{saddle}
\Gamma = \gamma^2 \sum_{n,p} S_{\Gamma}(\delta E_{n,p}) \exp(-2(n+p)/\Xi) 
\end{equation}
where we sum over all processes that involve an $n$ particle rearrangement in the system and a $p$ particle rearrangement in the bath, assuming that the linewidth is the same for all spectral lines, and where $\delta E_{n,p} \simeq \delta_m K^{-(m+n)}$ is the typical detuning from the nearest spectral line at that order. 
In the above expression we have assumed that the dominant contribution to the decay rate comes from processes at leading order in $\gamma$. This approximation is valid when $\ln(\delta_m/\gamma) \gg 1/\Xi$, so that it is less costly to go to high order in the system or in the bath than to go to high order in the system-bath coupling~\footnote{Taking into account the definition of $\Xi = 1/\ln(g \delta_m / \lambda \Delta)$, we see that the condition for validity of the above leading order in $\gamma$ approximation is $
\frac{g \gamma}{\lambda \Delta} \ll 1$,
i.e., the inter-dot coupling $\gamma$ must be weak enough}. Assuming that this condition is satisfied, the equation (\ref{saddle}) can be rewritten explicitly as

\beq\label{selfconbig}
\Gamma = \frac{\gamma^2 \Gamma}{\delta_m^2} \sum_{n,p} \frac{ e^{-2(n+p)(1/\Xi - \ln K)} n_{fast}(K^{n+p}/\delta_m)}{1 + [\Gamma n_{fast}(K^{n+p}/\delta_m) K^{n+p} / \delta_m]^2}
\eeq
where
\beq\label{nfast'}
n_{fast}(t) = \left\{\begin{array}{cc} 1 & t \ll 1/\gamma \\ s(T) N \ln^d(\gamma t) & 1/\gamma \ll t \ll 1/\Gamma \\ s(T) N  \ln^d(\gamma/\Gamma) & t \gg 1/ \Gamma\end{array} \right.
\eeq
This pair of equations is the central result of our self-consistent mean-field theory. In what follows, we shall explore its behavior both numerically and, in some limits, using saddle-point methods.

\section{Mean-field solutions}

\subsection{Numerical solution and limiting cases}

The sum in Eq.~\eqref{selfconbig} can be approximated as an integral over $n+p$, which can then be solved self-consistently for $\Gamma$. One expects this approximation to be good whenever the dominant values of $n+p$ are large, i.e., whenever rearrangements are large and collective. (Moreover, this approach agrees well with that of directly evaluating the sum (see Appendix), wherever the mean-field equations are applicable.)

Before we turn to the numerical solution, let us briefly discuss some limiting cases. First, when $\Xi \log K \ll 1$, the sum in Eq.~\eqref{selfconbig} is dominated by its first term. One can check that in this limit there are no self-consistent solutions for $\gamma \ll \delta_m$. Second, when $\Xi \log K$ is greater than and not too close to 1, the sum can be evaluated by saddle-point methods (see Appendix C-E) yielding the dependence, 

\begin{equation}\label{agklscaling}
\Gamma = \gamma^{\frac{\Xi \ln K}{\Xi \ln K -1}}
\end{equation}
Note that this scaling is identical to that for a single dot.

\begin{figure}[h]
\begin{center}
\includegraphics[width=0.4 \textwidth]{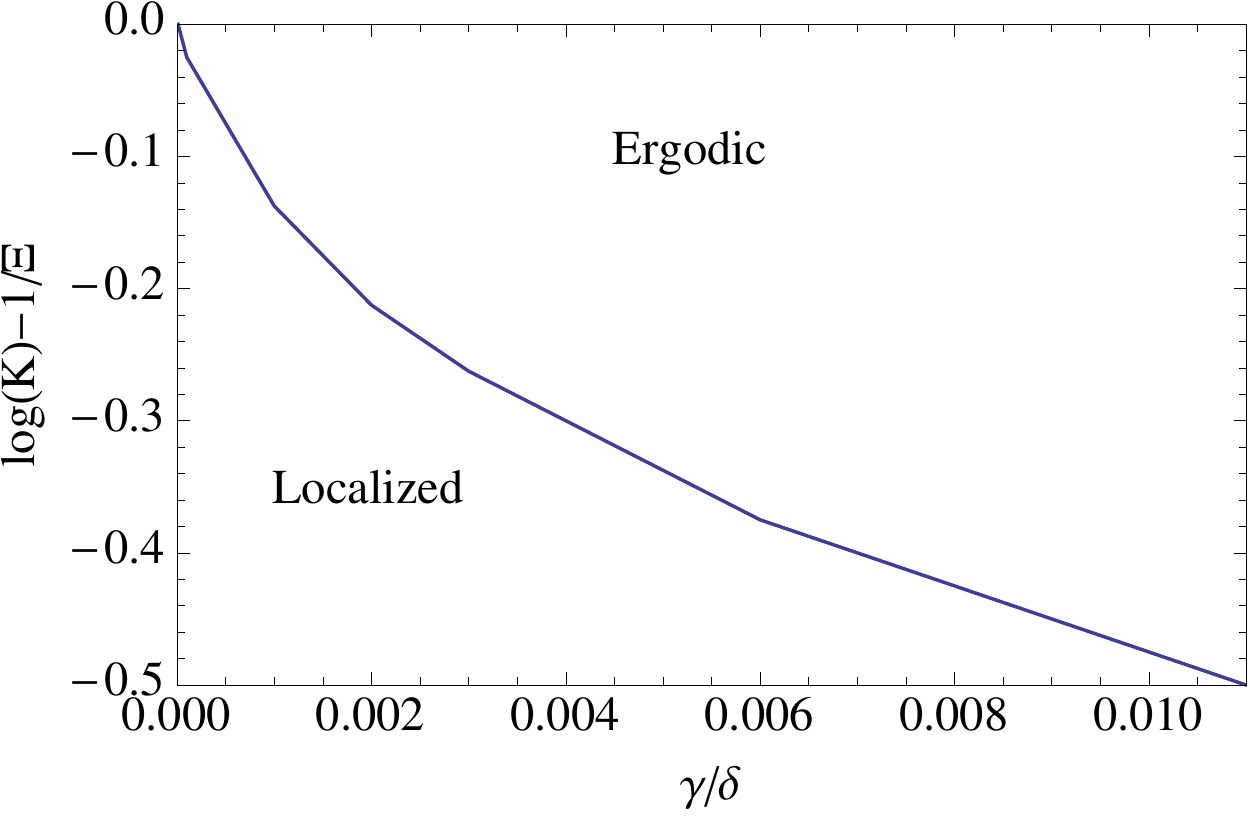}
\caption{Numerically computed mean-field phase diagram. The ergodic phase is stabilized, even in the regime where a single dot would be localized, by the inter-dot coupling. The phase boundary is approximately linear (i.e., $\gamma_c/\delta_m \sim 1/\Xi - \log K$) for the larger values of $\gamma$, but approximately logarithmic (i.e., $\log(\gamma_c/\delta_m) \sim \log K - 1/\Xi$) for smaller $\gamma$.}
\label{phasediagram}
\end{center}
\end{figure}

\begin{figure}[h]
\begin{center}
\includegraphics[width=0.4 \textwidth]{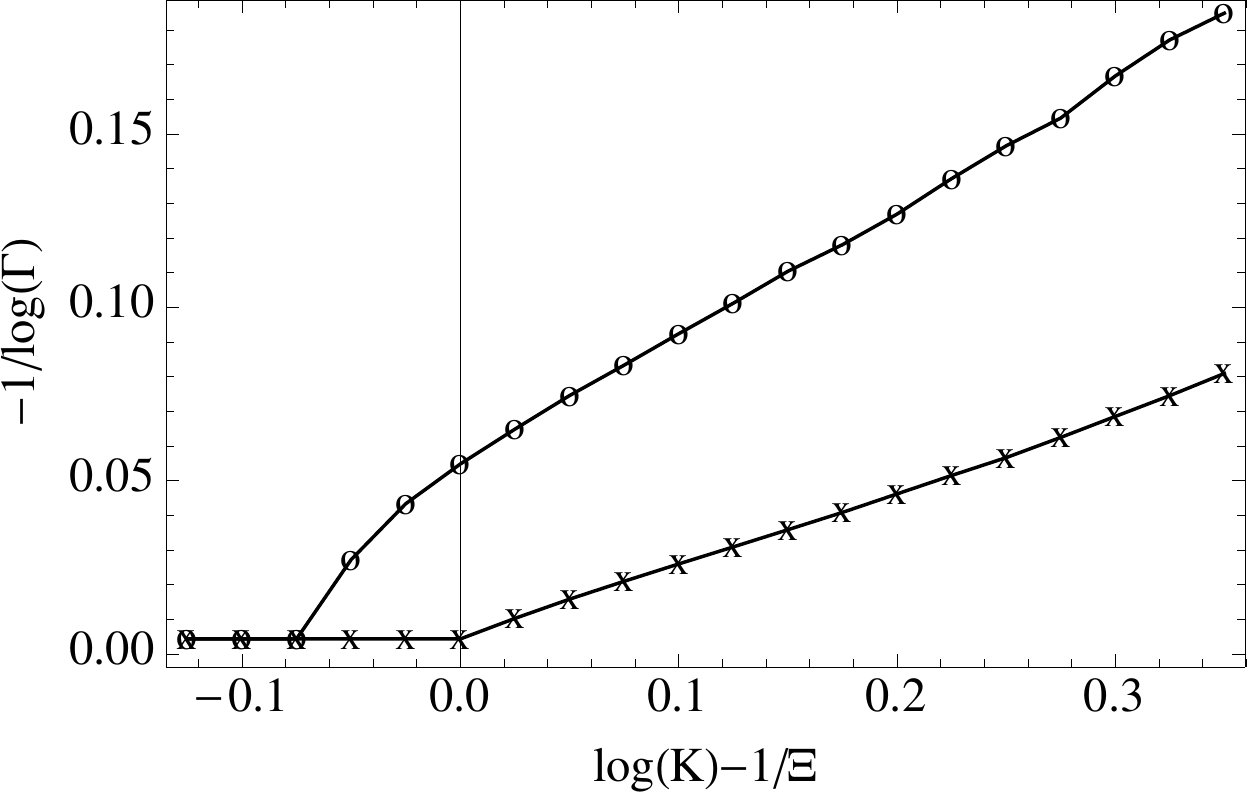}
\caption{Numerical solutions of the mean-field equations for $N = 10, K = 3$, and $\gamma/\delta_m = 10^{-4}$ (crosses) and $\gamma/\delta_m = 10^{-2}$ (circles). In the saddle-point approximation $-1/\log(\Gamma)$ should vanish linearly at $\Xi \log K = 1$ (i.e., at the origin). As the inter-dot coupling is decreased, the saddle-point equations become more accurate. The parameters $\Xi$ and $K$ are defined in the main text.}
\label{agklonset}
\end{center}
\end{figure}

\begin{figure}[h]
\begin{center}
\includegraphics[width=0.4 \textwidth]{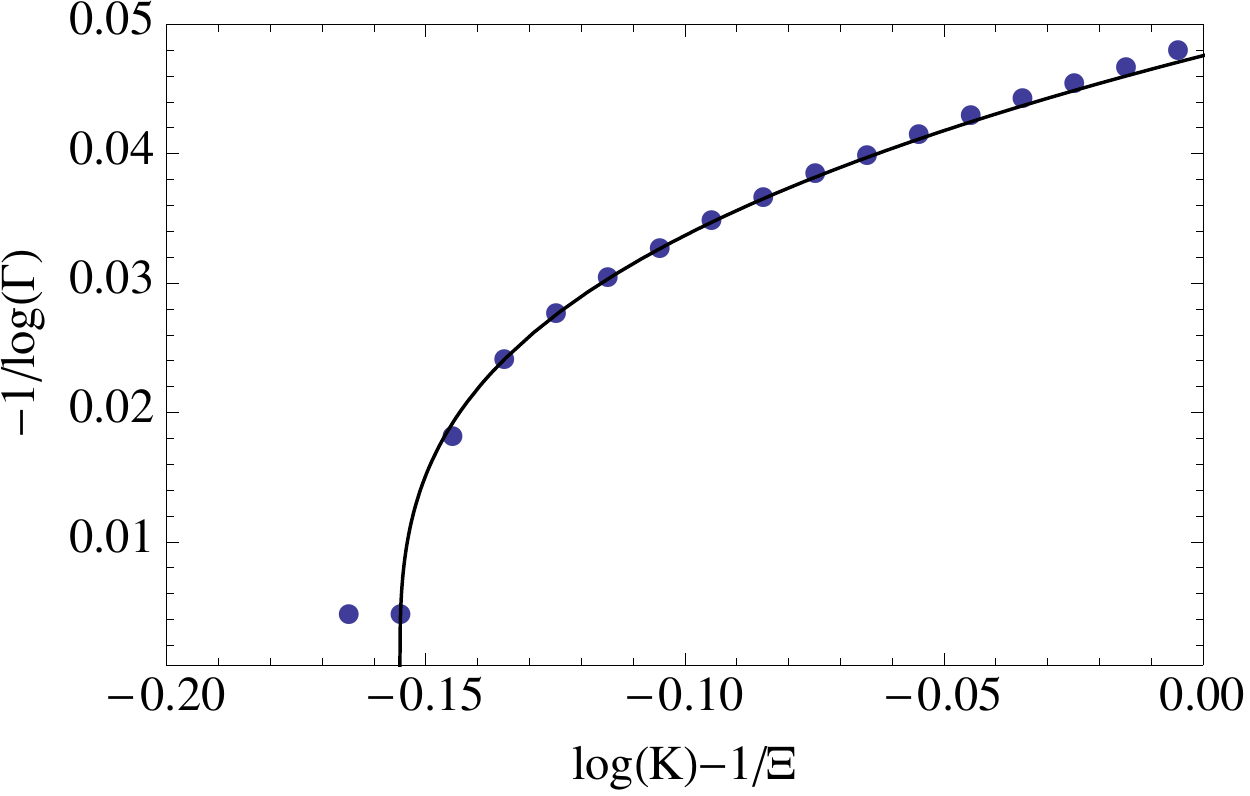}
\caption{Numerical solutions of the mean-field equations for $N = 10, K = 10$, and $\gamma/\delta_m = 10^{-3}$ showing how the linewidth goes to zero. The data fits well to the critical behavior $-1/\log(\Gamma) \sim (\Xi - \Xi_c)^{1/3}$.}
\label{critical}
\end{center}
\end{figure}

We now turn to our numerical results, which are presented in Figs.~\ref{phasediagram}, \ref{agklonset}, and \ref{critical}. Fig.~\ref{phasediagram} shows the mean-field phase diagram, which is controlled by the Cayley-tree parameter $\Xi \log K - 1$. When this parameter is positive, each quantum dot is susceptible to delocalizing (and would in fact be delocalized if it were infinitely large), so any coupling between the dots gives rise to a metallic state. By contrast, when this parameter is negative, the states on an individual dot are localized, and the inter-dot coupling must therefore be above a critical value in order to establish a metal. This critical value increases as $\Xi$ decreases; eventually $\gamma_c$ becomes of order $\delta_m$ and the mean-field procedure ceases to be consistent.

Figs.~\ref{agklonset} and \ref{critical} show how the linewidth vanishes near the MBL transition. Fig.~\ref{agklonset} shows the vanishing of the linewidth both for very weak inter-dot coupling $\gamma$ and for stronger coupling. Consistent with our intuition, as $\gamma/\delta \rightarrow 0$ we recover single-dot scaling, as the system increasingly behaves like a set of independent dots. However, even for very small $\gamma/\delta$ one sees substantial deviations from single-dot scaling. Finally, Fig.~\ref{critical} zooms in on the ``critical'' region where the linewidth vanishes. Within our mean-field theory, we find that the linewidth vanishes as $-1/\ln(\Gamma) \sim (\Xi - \Xi_c)^{1/3}$, or equivalently 

\beq
\Gamma \sim \exp\left( - \frac{\lambda}{(\Xi - \Xi_c)^{1/3}} \right).
\eeq
with some nonuniversal prefactor $\lambda$. 

\subsection{Regime of validity of mean field equations}
In deriving the self consistent mean field equations (\ref{selfconbig}, \ref{nfast'}) we made several assumptions. We now catalog the various assumptions made, and discuss the regime of validity of the resulting equations. In two separate places, we assumed that $\gamma$ was `weak enough.' Firstly, we assumed that $\gamma/\delta_m \ll 1/z$ (where $z$ is the coordination number) so that the dominant spectral lines do not overlap. Next, we assumed $\gamma/\delta_m \ll \exp(-1/\Xi)$, to justify restricting ourselves to lowest order in $\gamma$. These two assumptions can be combined into the condition 
\begin{equation}
\frac{\gamma}{\delta_m} \ll \min \left( \exp(-1/\Xi), 1/z \right) \label{condition1}
\end{equation}
Additionally, to justify application of the Golden Rule, we required (Appendix B) that 
\begin{equation}
\frac{\gamma}{\delta_m} \gg \sqrt{z N^2} K^{-N} \label{condition2}
\end{equation}
where $N$ is the number of single particle states on a dot. We thus conclude that our derivation of the mean field equations is valid over a parameter range
\begin{equation}
\sqrt{z N^2} K^{-N} \ll \frac{\gamma}{\delta_m} \ll \min \left( \exp(-1/\Xi), 1/z \right) \label{range}
\end{equation}
This regime of validity can be made broad by taking $N$ to be large. Finally, in order to justify working with typical rather than average bath spectra (Appendix B), we assumed 
\begin{equation}
\frac{\Gamma}{\gamma} \ln^{2d} \frac{\gamma}{\Gamma} \ll 1 \label{keycondition'}
\end{equation}
Additionally, our approach is only sensible if the decay rate obtained is larger than the many body level spacing of the system i.e. $\Gamma > N \delta_m K^{-N}$. Combining this condition with Eq.\ref{keycondition'} we conclude that the mean field equations apply IFF
\begin{equation}
N K^{-N} \ll \frac{\Gamma}{\delta_m} \ll \frac{\gamma}{\delta_m} \label{keycondition}
\end{equation}
Thus, the mean field approach becomes inconsistent when we get too close to the critical point (so that the line broadening is less than the many body level spacing on the dot), and also when we got too far from the critical point (when the central assumption of an inhomogenous bath spectrum breaks down). However, Eq.\ref{range} guarantees that there is a non-vanishing regime in the vicinity of the critical point where the mean field equations can be safely applied. This `window of applicability' may be extended arbitrarily close to the critical point by making $N$ large. 

\subsection{Temperature-dependence of the linewidth and conductivity}
Thus far we have treated $\Xi$, the Fock space localization length on a dot, as our tuning parameter. In practice, $\Xi$ is related to the disorder strength, the temperature, and the interaction strength. Of these quantities, the temperature is the most natural tuning parameter in experiment; therefore, we close this section by re-expressing our main results in terms of the temperature $T$. We note that $T$ enters into our expressions via three quantities~\cite{agkl}: the entropy density $s(T)$, which is $\propto T$ at low temperatures and essentially constant at high temperatures; the three-particle level spacing $\delta_m(T) \sim \Delta^3/T^2$; and the Fock-space localization length $\Xi \approx 1/\ln[g \delta_m(T)/(\lambda \Delta)]$. As one might expect, $\Xi$ decreases with temperature.

Substituting these relations into Eq.~\eqref{solution}, we find%
\begin{equation}
\Gamma(T>T_0) \sim \bigg(\frac{\gamma T^{5/2} }{\Delta^3} \bigg)^{T_0 \ln K/2(T-T_0)}. \label{scalingsolution}
\end{equation}
We recall \cite{ngh} that the conductivity scales as $\sigma \sim \Gamma$, and thus the conductivity should scale the same way within the regime of validity of the saddle point solution. However, the saddle point approximation is only well controlled as long as we satisfy Eq.\ref{regimeofvalidity}, which in the temperature regime means 
\begin{equation}
\frac{T-T_0}{T_0} > \frac{1}{4}
\end{equation}
Once we get any closer to $T_0$ than this bound, we can no longer use the saddle point solution, but instead must consider the full numerical solution of the mean field equations. We note that a fit to the numerical solutions suggests that the near critical scaling follows $\Gamma \sim \exp(-1/(\Xi - \Xi_c)^{1/3})$. This would in turn imply a temperature scaling
\begin{equation}
\sigma \sim \exp(-a''/(T-T_0')^{1/3})
\end{equation}
However, since this near critical regime is analytically intractable, we cannot be certain that this scaling form is truly universal. 

\section{Conclusions}
We have argued that a many-body localized system coupled to a slowly fluctuating external bath exhibits rich relaxational dynamics due to collective hops; in particular, the relaxation time diverges as a power law of the bandwidth of the bath. We have argued that our results apply to a wide range of realistic experiments. 
The slow relaxational dynamics due to an extrinsic (e.g., nuclear-spin) bath is intimately related to the slowing of relaxation as the MBL transition is approached; thus, as we have discussed, the narrow-band relaxation results can be used to construct a self-consistent mean-field description of a metal near the MBL transition (i.e., in the regime where the intrinsic relaxation timescales are parametrically slower than the characteristic dynamical timescales of the system). 

The key quantity that we examine as a diagnostic of proximity to the localized phase is the typical width of a local spectral line; this quantity, which is related to observables such as the DC conductivity, would be strictly zero in the localized phase. As the localized phase is approached, the line width scales to zero first as Eq.~\eqref{scalingsolution}, and then more slowly (in a manner that can be captured by numerical solution of a self consistency equation). We have also estimated how the DC conductivity should scale to zero as the localized phase is approached, and predict that it should originally follow Eq.~\eqref{scaling}, before crossing over to a slower behavior at the lowest temperatures (Fig.4). We note that this expectation appears consistent with experimental data reported in \cite{Shahar}. 

We have assumed that a finite-linewidth state is ergodic, there is some numerical evidence~\cite{randomgraph} for an intermediate phase that is conducting but \emph{not} ergodic, in which the eigenstates have sub-extensive entanglement entropy~\cite{grover}. Our mean-field theory cannot directly distinguish such a phase from an ergodic metal, though it is possible that with some refinements (e.g., treating distributions of linewidths~\cite{Fiegelman}, or using the entanglement entropy density instead of the thermal entropy density) it might be adapted to describe such an intermediate phase. 

Our mean field approach does also assume that the system is spatially homogenous. This assumption of spatial homogeneity is expected to be a poor approximation to the true physics in the vicinity of the critical point. However, a generalization of our approach that (numerically) treats spatially inhomogenous systems appears (at least conceptually) straightforward.

To summarize, our mean-field approach, which consists of treating a correlation volume as a zero-dimensional, interacting system coupled to a self-consistent bath, should be generalizable to a wide range of models with MBL transitions. Moreover, our predicted scaling of the DC conductivity with temperature as one approaches the many body localization transition should be directly testable in experiments, and appears consistent with recent results \cite{Shahar}. 

\emph{Acknowledgments}. We are indebted to David Huse, Ariel Amir, Michael Knap, Norman Yao, Eugene Demler, and Mikhail Lukin for helpful discussions; and to David Huse and Ariel Amir for a critical reading of this manuscript. This research was supported in part by the Harvard Quantum Optics Center (SG) and by a PCTS fellowship (RN), and began at the Boulder summer school for condensed matter physics.


%


\appendix

\section{Gaussian and Lorentzian baths}

The arguments in the main text were framed in terms of a bath with a constant density of states over a window $[-W/2,W/2]$. We now generalize these results to baths with spectra that are Gaussian (e.g., a nuclear spin-bath) and Lorentzian (e.g., a self-consistent bath). For concreteness, we shall focus on the relaxation rate Eq.~\eqref{manyLbit}, but the generalization to the other main results in the main text is straightforward. 
\subsection*{Gaussian bath}

First, we consider the case of a bath with the density of states $N(E) \sim \exp[-(E/W)^2]/W$. Applying the Golden Rule, we find that the relaxation rate of an $n^d$-particle rearrangement goes as

\beq
\Gamma_n \sim \frac{\gamma^2}{W} \exp\left(-\frac{2n^d}{\Xi} - \frac{\Delta^2}{W^2} e^{-2 s(T) n^d} \right).
\eeq
One can optimize $\Gamma_n$ by steepest-descent methods. The solution is:

\beq
\hat{n}^d = \frac{1}{2s(T)} \ln\left(\frac{\Delta^2}{W^2} \frac{\Xi s(T)}{2} \right)
\eeq
which agrees, up to logarithmic corrections, with Eq.~\eqref{manyLbit}. 

\subsection*{Lorentzian bath}

As discussed in the main text surrounding Eq.~(16), the case of a Lorentzian bath has two distinct regimes, depending on whether the matrix element decreases faster or slower than the density of states increases as we go to higher order in Fock space. One can see this as follows: for a decay process that uses the tails of the Lorentzian (of width $W$), the Golden-Rule decay rate can schematically be written as $W ($matrix element/energy denominator$)^2$. The quantity in parentheses is precisely the perturbative expression for the wavefunction amplitude on the target site; therefore, this quantity (and the Golden-Rule decay rate) must decay with $n$ whenever the wavefunction is localized. To state this more quantitatively: for a density of states $N(E) \sim W/(E^2 + W^2)$, the typical matrix element for an $n$th order process is suppressed as $e^{-n/\Xi}$. By contrast, the level spacing scales as $\exp(-s(T)n)$. Localization at a given temperature requires that $s(T) \Xi < 1$.

The regime of collective relaxation for a Lorentzian bath only exists (on a Cayley tree) when $\Xi \ln K > 1$. Here, the relaxation rate of an $n$-step hop goes as

\beq
\Gamma_n \sim \frac{\gamma^2}{W} \frac{e^{-2 n/\Xi}}{(\Delta/W)^2 e^{-2 n \ln K} + 1}
\eeq
Optimizing with respect to $n$ yields the condition

\beq
e^{- 2 n \ln K} = \frac{W^2}{\Delta^2} \frac{1/\Xi}{\ln K - 1/\Xi},
\eeq
which leads to the solution

\beq
\hat{n} = \frac{1}{\ln K} \left[\ln(\Delta/W) + \frac{1}{2} \ln(\Xi \ln K - 1) \right]
\eeq
which is again the same form as that in the main text. 

Finally we comment on the case in which the wings of the Lorentzian have a hard (or Gaussian) cutoff at an energy $\Lambda$. In this case, the lowest-order process that the bath can mediate is at the order $n_0 \sim \ln(\Delta/\Lambda)/\ln K$. When $\Xi \ln K < 1$, one can conclude from the arguments above that relaxation takes place at exactly this order, and crucially, that the optimal order in perturbation theory is \emph{independent} of $W$. 

\section{Validity of the Golden Rule}
In this appendix we justify our use of the Golden Rule in estimating the decay rate of the coupled dot problem. There are three potential problems with the use of the Golden Rule (none of which, however, is important in the regime of interest). 

Firstly, when $\gamma/\sqrt{zN^2} < \delta_m K^{-N}$, then the matrix element between dots is less than the typical level spacing on a finite size dot. In this limit, the Golden Rule is inapplicable and the decay rate is strictly zero. However, we are working in the large $N$ limit when $\gamma/\sqrt{z N^2} \gg \delta_m K^{-N}$. 

Secondly, for small enough $\Gamma$, we expect that the matrix-element for system-bath coupling $\gamma/\sqrt{zN^2} \gg \Gamma$, so that in principle the ``back-action'' of the system on the bath might be important. Physically, this back-action corresponds to the fact that the bath energy levels fluctuate whenever the system undergoes a transition; this effect is precisely what gives rise to the ``Hartree'' broadening of bath levels, which we discussed. Thus, our approach has already incorporated back-action at the mean-field level, which is appropriate in the limit of large coordination number---in this limit, most of the fluctuations of a given bath level are due to \emph{other} bath levels, rather than due to the system. 

The final difficulty is that the system and bath levels are \emph{correlated} as a result of local level repulsion. Thus, in the MBL phase, the bath would have \emph{no} spectral weight at the system transition frequency. If one ignores these correlations altogether, and uses \emph{averaged} bath spectra, one incorrectly finds that there is no localization transition, even in the single-dot case~\cite{agkl}. However, this difficulty can be addressed by working with \emph{typical} rather than average bath spectra: in the typical spectra, rare resonances between system and bath levels are suppressed, and the physics of localization transitions is captured~\cite{dobr}. In the many-body case, the consistency criterion for using typical rather than average spectra is that the bath spectral lines do not self-average (by diffusing through their entire spectral range) on the timescale of a typical system transition; i.e., we need the condition $\Gamma^{-1} \ll \gamma/(\Gamma \ln^d(\gamma/\Gamma))^2$. This condition is indeed satisfied near the MBL transition, when $\Gamma \rightarrow 0$.

\section{Saddle point solutions to the self consistency equations}
There are various different regimes, and we discuss each in turn. 
\subsection{The localized phase $\frac{1}{\Xi} \gg \ln K$}
When $1/\Xi \gg \ln K$, the dominant rearrangements are those at lowest order ($n,p \approx 1$). Moreover, we are assuming that $\delta_m /K^2 \gg \gamma$ (i.e. we have weakly coupled dots), so $n_{fast} \approx 1$ for the dominant processes. If we further demand that the decay rate $\Gamma$ should be real, then we discover that the only self consistent saddle point solution to (18, 19) in the regime $1/\Xi \gg \ln K$ and $\delta_m /K^2 \gg \gamma$ is $\Gamma = 0$ i.e. this regime corresponds to the localized phase. 

\subsection{The delocalized phase $\frac{1}{\Xi} < \ln K$}
When $\frac{1}{\Xi} < \ln K$, then the numerator of Eq.\ref{selfconbig} is an increasing function of $n$ and $p$. Meanwhile, the denominator is of order one for small $n+p$, but cuts off the summand for $n+p > N_c$, where $N_c$ is set by the condition 
\begin{equation}
\label{nc}
(\Gamma/\delta_m) K^{N_c} n_{fast}(K^{N_c}/\delta_m) = 1. 
\end{equation}
An explicit solution for $N_c$ is obtained in Appendix D. We find that 
\begin{equation}
N_c \ln K = \ln(\delta_m/\gamma) + d \mathcal{W}\left(\frac{(\gamma/(\Gamma s(T) N))^{1/d}}{d}  \right) \label{nc1}
\end{equation}
where $\mathcal{W}$ is the Lambert W function. In the limit of small $\Gamma$ (strictly the limit $\Gamma s(T) N \ll \gamma$), this asymptotes to 
\begin{equation}
\label{nc2}
N_c \ln K \approx \ln(\delta_m/\Gamma s (T) N d^d).
\end{equation}
 Comparing with Eq.\ref{nc}, we conclude that in the limit $\Gamma \rightarrow 0$, $n_{fast}(K^{N_c}/\delta_m)$ saturates to a $\Gamma$ independent value $n_c \approx s(T) N d^d$. 

In the regime $1/ \Xi < \ln K$, when the numerator of Eq.\ref{selfconbig} is an increasing function of $n+p$, the sum in Eq.\ref{selfconbig} is dominated (at the saddle point level) by the largest terms we can get before the growth of the denominator cuts off the summand i.e. the sum is dominated by the $\approx N_c$ terms with $n+p \approx N_c$. Making use of the defining relation Eq.\ref{nc} we can approximate Eq.\ref{selfconbig} by 
\begin{equation}
\label{sp}
\frac{\Gamma}{\delta_m} = \frac{\gamma^2}{\delta_m^2} N_c K^{N_c} \exp\left( - \frac{2 N_c}{\Xi} \right)
\end{equation}
This equation, together with Eq.\ref{nc1}, in principle fully determines $\Gamma/\delta_m$ in terms of the parameters $s(T) N, \Xi, \gamma/\delta_m$ and $K$. However, to obtain a closed form solution we make use of the $\Gamma \rightarrow 0$ asymptotic form for Eq.\ref{nc1} i.e. Eq.\ref{nc2}. Substituting Eq.\ref{nc2} into Eq.\ref{sp}, taking logs, dropping $\ln(\ln(\delta/\Gamma))$ corrections (which have already been ignored in approximating Eq.\ref{nc1} by Eq.\ref{nc2}), we find that (with logarithmic accuracy)
\begin{equation}
\Gamma = \frac{\delta}{n_c} \left(\frac{\gamma \sqrt{n_c}}{\delta}\right)^{\frac{\Xi \ln K}{\Xi \ln K -1}}; \qquad n_c \approx s(T) N d^d \label{solution}
\end{equation}
This is a similar scaling behavior to that obtained in the single quantum dot. Thus, we find that the saddle point analysis essentially reproduces the `single quantum dot' scaling behavior, with the main correction coming through the pre factor $n_c$. In simplifying Eq.\ref{nc1} to Eq.\ref{nc2}, and in proceeding from Eq.\ref{sp} to Eq.\ref{solution}, we retained terms in $\ln(\delta/\Gamma)$ but dropped terms in $\ln \ln (\delta/\Gamma)$. This `leading log' approximation becomes increasingly good as the critical point is approached. 
\subsection{Regime of validity of saddle point solutions}
Firstly, we recall that self consistency requires that the saddle point solution Eq.\ref{solution} satisfy the inequality Eq.\ref{keycondition}. This requires that (upto log corrections)
\begin{equation}
\frac{\gamma^2}{\delta^2} s(T) N d^d< 1;  \quad \frac{\ln(\delta_m/\gamma s(T) N)}{N \ln K} < \frac{\Xi \ln K -1}{\Xi \ln K } < 1 \label{sc1}
\end{equation}
The first condition (taken in conjunction with Eq.\ref{condition1},\ref{condition2}), constrains our theory: if we take the limit of large $N$ to satisfy Eq.\ref{condition2}, we must simultaneously scale $\gamma$ as $N^{-1/2}$ in order for our derivations to hold. The second condition simply tells us that our approach only works in a window where we are close enough to the transition for the bath spectrum to be highly inhomogenous, but not so close that the self consistently determined line broadening is less than the many body level spacing on the dot. 

 \begin{figure}
 \includegraphics[width=\columnwidth]{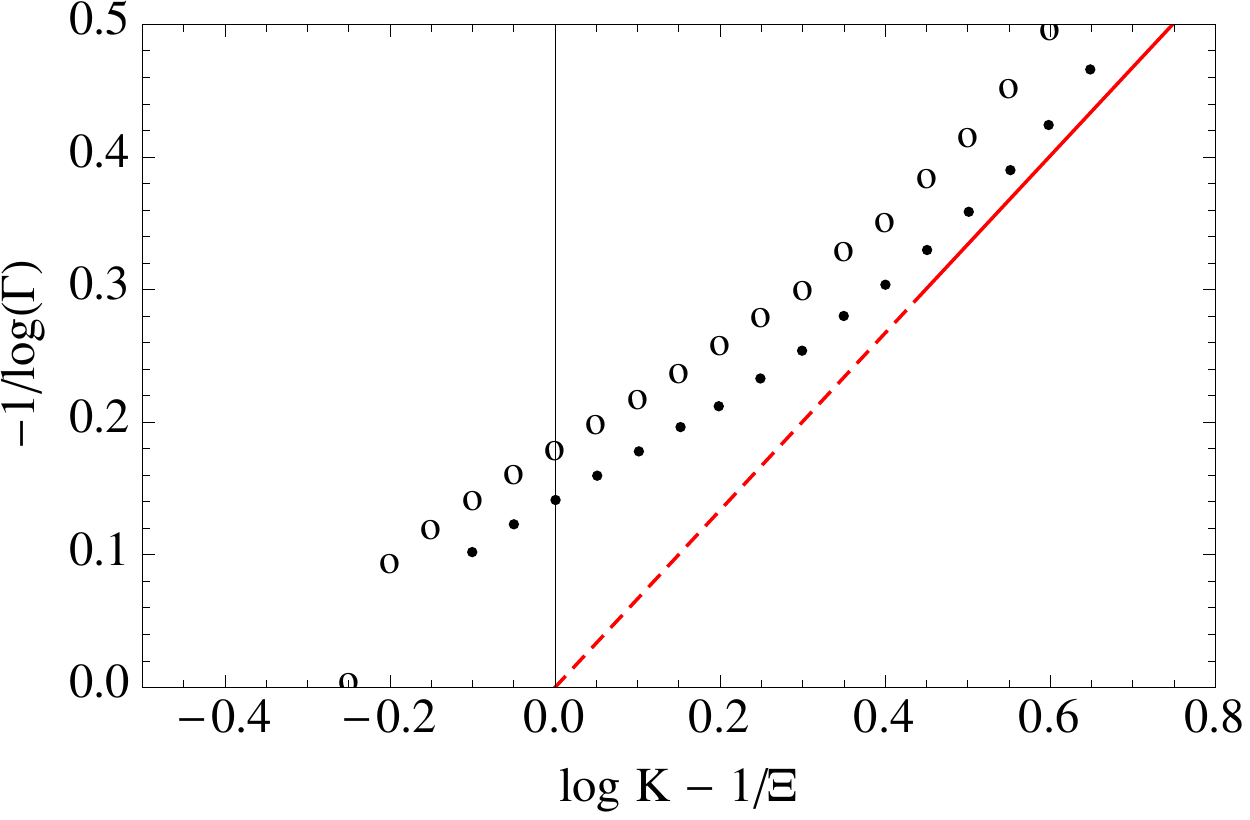}
 \caption{\label{fig: data} Plot showing how $-1/\ln (n_c \Gamma)$ scales with $\epsilon$. The black data points are obtained by numerical solution of the equation Eq.\ref{approxeq}, for parameters $\gamma = 0.1$, $s =0.5$, $N=10$, $d=1$ and $K=3$ (a parameter choice that yields $n_c = 5$). Open circles are obtained by approximating the mean-field Eq.~\eqref{selfconbig} with an integral; the two numerical approaches predict the same behavior. 
The red curve is the theory prediction Eq.\ref{solution} (solid line in the range of validity of the saddle point approximation, dashed outside the range of validity). We emphasize that the red curve is exactly the saddle point solution---there are no fitting parameters.  We can see that Eq.~\eqref{solution} provides a reasonable approximation to the full numerical solution (data points) within the regime of validity of the saddle point approximation $0.45< \epsilon < 0.8$. However, for $\epsilon < 0.45$ it dramatically underestimates $\Gamma$, which goes to zero much more slowly than would be predicted by a scaling of the form Eq.~\eqref{solution}. }
 \end{figure}

Finally, the saddle point approximation itself is controlled by the value of $\ln K - \frac{1}{\Xi}$. For $ \ln K - \frac{1}{\Xi} > 1/2$, the sum Eq.\ref{selfconbig} is well approximated by the saddle point Eq.\ref{sp}. However, for $\ln K - \frac{1}{\Xi} < 1/2$, the saddle point Eq.\ref{sp} provides a poor approximation to the sum, as terms with $n+p \neq N_c$ also make a substantial contribution. Combining this with Eq.\ref{sc1}, we conclude that the saddle point solution Eq.\ref{solution} well approximates the behavior over the parameter regime
\begin{equation}
\frac{1}{2\ln K} < \frac{\Xi \ln K - 1}{\Xi \ln K} < 1 \label{regimeofvalidity}.
\end{equation}
If we are too far from the critical point, the assumptions leading to the self consistency equations break down. If we are too close to the critical point, the self consistency equations themselves are fine, but the saddle point approximation becomes increasingly unreliable, and the equations must be solved numerically. There is however a substantial intermediate regime where the saddle point solution Eq.\ref{solution} is reliable, and this window can be extended arbitrarily close to the critical point by taking $K \rightarrow \infty$. 

\section{Solving for the saddle point parameter $N_c$}
We recall that $N_c$ is defined by 
\begin{equation}
K^{N_c} \frac{\Gamma}{\delta_m} n_{fast}(K^{N_c}/\delta_m) = 1
\end{equation}
Now $n_{fast}$ lies in the range $1 \le n_{fast} \le s(T) N \ln^d(\gamma/\Gamma)$. This then implies that $N_c$ must lie in the range 
\begin{equation}
\frac{\ln \frac{\delta_m}{\Gamma s(T) N \ln^d(\gamma/\Gamma) }}{\ln K}  \le N_c \le \frac{\ln \delta_m / \Gamma}{\ln K}
\end{equation}
We saturate the upper bound IFF $n_{fast} = 1$. However, to obtain $n_{fast} = 1$ we require that $\gamma K^{N_c}/\delta _m< 1$ i.e. $N_c =  \frac{\ln \delta_m / \Gamma}{\ln K} < \ln(\delta_m / \gamma)/\ln K$. To satisfy this self consistency condition  requires assuming that $\Gamma > \gamma$ i.e. the relaxation rate due to inter dot coupling is bigger than the inter dot coupling itself. This solution is self evidently unphysical, and should be discarded. Meanwhile, the lower bound solution for $N_c$ is obtained if we take $n_{fast} = s(T) N \ln^d(\gamma/\Gamma)$, but for this to be consistent we require that $\Gamma K^{N_c} /\delta_m > 1$ i.e. $N_c = \frac{\ln \frac{\delta_m}{\Gamma s(T) N \ln^d(\gamma/\Gamma) }}{\ln K} > \ln(\delta_m/\Gamma)/\ln K$. This inequality is clearly violated, so a solution at the lower bound in (D2) is inconsistent. A consistent and physically reasonable solution for $N_c$ must thus lie between tighter bounds than those presented in (D2), and must have $n_{fast}$ given by the `intermediate time' form $n_{fast} \approx s(T) N \ln^d (\gamma K^{N_c}/\delta_m)$. Thus we conclude that $N_c$ is the solution to the equation
\begin{equation}
N_c = \frac{\ln\left[\frac{\delta_m}{\Gamma s(T) N \ln^d(\gamma K^{N_c}/\delta_m) } \right]}{\ln K}
\end{equation}
This equation may be solved exactly using MATHEMATICA to yield
\begin{equation}
N_c \ln K = \ln(\delta_m/\gamma) + d \mathcal{W}\left(\frac{(\gamma/(\Gamma s(T) N))^{1/d}}{d}  \right) 
\end{equation}
where $\mathcal{W}$ is the Lambert W function. In the limit of small $\Gamma$ (strictly the limit $\Gamma s(T) N \ll \gamma$), this asymptotes to $N_c \ln K \approx \ln(\delta_m/\Gamma s (T) N d^d)$. Comparing with Eq.D1, we conclude that in the limit $\Gamma \rightarrow 0$, $n_{fast}(K^{N_c}/\delta_m)$ saturates to a $\Gamma$ independent value $n_c \approx s(T) N d^d$. (This level of approximation essentially amounts to dropping the $\ln(\ln)$ correction in D3).

\section{Direct numerical solution of mean field equations}
In the main text we presented numerical solutions of Eq.~\ref{selfconbig} that were attained by approximating the sum by an integral. In this section, we present a different numerical solution of the mean field equations, as a function of the parameters $d, \gamma/\delta_m, K, N$ and $s(T)$. The solutions are obtained for $d=1$, $\gamma/\delta_m = 0.1$, $K = 3$, $N = 10$ and $s = 0.5$, a choice that respects the various inequalities discussed in Section IVE. We then numerically solve for $\Gamma$ as a function of $\frac{1}{\Xi} - \ln K$. 

We begin by noting that (as discussed in Appendix C), the sum in Eq.\ref{selfconbig} is cut off by the growth of the denominator well before $n_{fast}$ saturates to its maximum value of $s N \ln^d(\gamma/\Gamma)$. Thus, in Eq.\ref{selfconbig} we can take $n_{fast} = 1$ for $n+p<N_0$, and $n_{fast} = s N \ln^d(\gamma K^{n+p}/\delta_m)$ otherwise, where $N_0 = (1+ s(T) N \ln (\delta_m/\gamma))/s(T) N \ln K$. Introducing the parameter $\epsilon = \frac{\Xi \ln K -1}{\Xi \ln K}$, and measuring all energies in units of $\delta_m$ (which we set to one), the self consistency equations then boil down to 
\begin{widetext}
\begin{equation}
\Gamma = \gamma^2 \Gamma \left[ \sum_{n+p<N_0} \frac{K^{2(n+p)\epsilon}}{1 + \Gamma^2 K^{2(n+p)}} + \sum_{n+p>N_0}^{n,p=N} \frac{s N K^{2(n+p)\epsilon} \ln^d (\gamma K^{n+p})}{1 + \Gamma^2 K^{2(n+p)} s^2 N^2 \ln^{2d} (\gamma K^{n+p}) }\right]; \quad N_0 = \frac{1+s N \ln(1/\gamma)}{s N \ln K}
\end{equation}
\end{widetext}
For the parameters we have chosen, we have $N_0 \approx 2$, and thus the equation above can be well approximated by
\begin{equation}
\Gamma = \gamma^2 \Gamma \sum_{n,p=1}^{N} \frac{s N K^{2(n+p)\epsilon} \ln^d (\gamma K^{n+p})}{1 + \Gamma^2 K^{2(n+p)} s^2 N^2 \ln^{2d} (\gamma K^{n+p}) }.\end{equation}
This is a self consistency equation with $\sim N^2$ terms. It is now useful to note that the terms in the sum depend only on $n+p$, and not on $n$ and $p$ individually. Moreover, there are $\sim n+p-1$ terms with a given value of $n+p$, and so the above equation can be rewritten as a self consistency equation with $\sim 2 N$ terms, which takes the form
\begin{equation}
\Gamma = \gamma^2 \Gamma \sum_{n=2}^{2N} (n-1) \frac{s N K^{2n\epsilon} \ln^d (\gamma K^{n})}{1 + \Gamma^2 K^{2n} s^2 N^2 \ln^{2d} (\gamma K^{n}) }.\label{approxeq} \end{equation}

We look for non-trivial solutions $\Gamma(\epsilon) \neq 0$.  We solve the above equation numerically on MATHEMATICA for $d=1, \gamma = 0.1, K=3, N=10$ and $s=0.5$, and for various values of $\epsilon$ in the range $0.01 < \epsilon < 0.85$. We stop at $\epsilon = 0.85$ because at this point the line broadening $\Gamma$ becomes comparable to $\gamma$, indicating that the mean field equations are no longer applicable (i.e. we violate the consistency condition Eq.23). Meanwhile, the condition Eq.\ref{regimeofvalidity} for our present choice of parameters indicates that the saddle point solution Eq.\ref{solution} should well approximate $\Gamma(\epsilon)$ as long as $\epsilon > 0.45$, but should become increasingly unreliable for smaller $\epsilon$.

 The numerical solution does appear to bear out these expectations. In the regime of validity of the saddle point equations $0.45 < \epsilon < 0.8$, the theoretical expression Eq.\ref{solution} well approximates the numerical solution of the saddle point equations. However, for $\epsilon < 0.45$ the theory expression Eq.\ref{solution} dramatically underestimates $\Gamma$ i.e. once we leave the regime of validity of the saddle point equations Eq.\ref{regimeofvalidity}, the line broadening $\Gamma$ decreases with $\epsilon$ much more slowly than would be predicted by Eq.\ref{solution}. This is illustrated by Fig.\ref{fig: data}. 
 

\begin{thebibliography}{10}%
\makeatletter
\providecommand \@ifxundefined [1]{%
 \ifx #1\undefined \expandafter \@firstoftwo
 \else \expandafter \@secondoftwo
\fi
}%
\providecommand \@ifnum [1]{%
 \ifnum #1\expandafter \@firstoftwo
 \else \expandafter \@secondoftwo
\fi
}%
\providecommand \enquote [1]{``#1''}%
\providecommand \bibnamefont  [1]{#1}%
\providecommand \bibfnamefont [1]{#1}%
\providecommand \citenamefont [1]{#1}%
\providecommand\href[0]{\@sanitize\@href}%
\providecommand\@href[1]{\endgroup\@@startlink{#1}\endgroup\@@href}%
\providecommand\@@href[1]{#1\@@endlink}%
\providecommand \@sanitize [0]{\begingroup\catcode`\&12\catcode`\#12\relax}%
\@ifxundefined \pdfoutput {\@firstoftwo}{%
 \@ifnum{\z@=\pdfoutput}{\@firstoftwo}{\@secondoftwo}%
}{%
 \providecommand\@@startlink[1]{\leavevmode}%
 \providecommand\@@endlink[0]{}%
}{%
 \providecommand\@@startlink[1]{%
  \leavevmode
  \pdfstartlink
   attr{/Border[0 0 1 ]/H/I/C[0 1 1]}%
   user{/Subtype/Link/A<</Type/Action/S/URI/URI(#1)>>}%
  \relax
 }%
 \providecommand\@@endlink[0]{\pdfendlink}%
}%
\providecommand \url  [0]{\begingroup\@sanitize \@url }%
\providecommand \@url [1]{\endgroup\@href {#1}{\urlprefix}}%
\providecommand \urlprefix [0]{URL }%
\providecommand \Eprint[0]{\href }%
\@ifxundefined \urlstyle {%
  \providecommand \doi [1]{doi:\discretionary{}{}{}#1}%
}{%
  \providecommand \doi [0]{doi:\discretionary{}{}{}\begingroup
  \urlstyle{rm}\Url }%
}%
\providecommand \doibase [0]{http://dx.doi.org/}%
\providecommand \Doi[1]{\href{\doibase#1}}%
\providecommand \bibAnnote [3]{%
  \BibitemShut{#1}%
  \begin{quotation}\noindent
    \textsc{Key:}\ #2\\\textsc{Annotation:}\ #3%
  \end{quotation}%
}%
\providecommand \bibAnnoteFile [2]{%
  \IfFileExists{#2}{\bibAnnote {#1} {#2} {\input{#2}}}{}%
}%
\providecommand \typeout [0]{\immediate \write \m@ne }%
\providecommand \selectlanguage [0]{\@gobble}%
\providecommand \bibinfo [0]{\@secondoftwo}%
\providecommand \bibfield [0]{\@secondoftwo}%
\providecommand \translation [1]{[#1]}%
\providecommand \BibitemOpen[0]{}%
\providecommand \bibitemStop [0]{}%
\providecommand \bibitemNoStop [0]{.\EOS\space}%
\providecommand \EOS [0]{\spacefactor3000\relax}%
\providecommand \BibitemShut [1]{\csname bibitem#1\endcsname}%
\bibitem{Shahar}%
  \BibitemOpen
  \bibfield{author}{%
  \bibinfo {author} {\bibfnamefont{M.}~\bibnamefont{Ovadia}} \emph{et~al.},\ }%
  \bibfield{journal}{%
  \bibinfo {journal} {arXiv:1406.7510}}%
   (\bibinfo {year} {2014})%
  \bibAnnoteFile{NoStop}{Shahar}%
\bibitem{anderson_absence_1958}%
  \BibitemOpen
  \bibfield{author}{%
  \bibinfo {author} {\bibfnamefont{P.~W.}\ \bibnamefont{Anderson}},\ }%
  \bibfield{journal}{%
  \Doi{10.1103/PhysRev.109.1492}{\bibinfo {journal} {Phys. Rev.}}\ }%
  \textbf{\bibinfo {volume} {109}},\ \bibinfo {pages} {1492} (\bibinfo {year}
  {1958})%
  \bibAnnoteFile{NoStop}{anderson_absence_1958}%
\bibitem{fleishman}%
  \BibitemOpen
  \bibfield{author}{%
  \bibinfo {author} {\bibfnamefont{L.}~\bibnamefont{Fleishman}}\ and\ \bibinfo
  {author} {\bibfnamefont{P.~W.}\ \bibnamefont{Anderson}},\ }%
  \bibfield{journal}{%
  \Doi{10.1103/PhysRevB.21.2366}{\bibinfo {journal} {Phys. Rev. B}}\ }%
  \textbf{\bibinfo {volume} {21}},\ \bibinfo {pages} {2366} (\bibinfo {year}
  {1980})%
  \bibAnnoteFile{NoStop}{fleishman}%
\bibitem{gornyi_interacting_2005}%
  \BibitemOpen
  \bibfield{author}{%
  \bibinfo {author} {\bibfnamefont{I.~V.}\ \bibnamefont{Gornyi}}, \bibinfo
  {author} {\bibfnamefont{A.~D.}\ \bibnamefont{Mirlin}},\ and\ \bibinfo
  {author} {\bibfnamefont{D.~G.}\ \bibnamefont{Polyakov}},\ }%
  \bibfield{journal}{%
  \Doi{10.1103/PhysRevLett.95.206603}{\bibinfo {journal} {Phys. Rev. Lett.}}\
  }%
  \textbf{\bibinfo {volume} {95}},\ \bibinfo {pages} {206603} (\bibinfo {year}
  {2005})%
  \bibAnnoteFile{NoStop}{gornyi_interacting_2005}%
\bibitem{basko_metalinsulator_2006}%
  \BibitemOpen
  \bibfield{author}{%
  \bibinfo {author} {\bibfnamefont{D.}~\bibnamefont{Basko}}, \bibinfo {author}
  {\bibfnamefont{I.}~\bibnamefont{Aleiner}},\ and\ \bibinfo {author}
  {\bibfnamefont{B.}~\bibnamefont{Altshuler}},\ }%
  \bibfield{journal}{%
  \Doi{10.1016/j.aop.2005.11.014}{\bibinfo {journal} {Ann. Phys.}}\ }%
  \textbf{\bibinfo {volume} {321}},\ \bibinfo {pages} {1126} (\bibinfo {year}
  {2006})%
  \bibAnnoteFile{NoStop}{basko_metalinsulator_2006}%
\bibitem{oganesyan_localization_2007}%
  \BibitemOpen
  \bibfield{author}{%
  \bibinfo {author} {\bibfnamefont{V.}~\bibnamefont{Oganesyan}}\ and\ \bibinfo
  {author} {\bibfnamefont{D.~A.}\ \bibnamefont{Huse}},\ }%
  \bibfield{journal}{%
  \Doi{10.1103/PhysRevB.75.155111}{\bibinfo {journal} {Phys. Rev. B}}\ }%
  \textbf{\bibinfo {volume} {75}},\ \bibinfo {pages} {155111} (\bibinfo {year}
  {2007})%
  \bibAnnoteFile{NoStop}{oganesyan_localization_2007}%
\bibitem{Znid}%
  \BibitemOpen
  \bibfield{author}{%
  \bibinfo {author} {\bibfnamefont{M.}~\bibnamefont{Znidaric}}, \bibinfo
  {author} {\bibfnamefont{T.}~\bibnamefont{Prosen}},\ and\ \bibinfo {author}
  {\bibfnamefont{P.}~\bibnamefont{Prelovsek}},\ }%
  \bibfield{journal}{%
  \bibinfo {journal} {Phys. Rev. B}\ }%
  \textbf{\bibinfo {volume} {77}},\ \bibinfo {pages} {064426} (\bibinfo {year}
  {2008})%
  \bibAnnoteFile{NoStop}{Znid}%
\bibitem{pal_many-body_2010}%
  \BibitemOpen
  \bibfield{author}{%
  \bibinfo {author} {\bibfnamefont{A.}~\bibnamefont{Pal}}\ and\ \bibinfo
  {author} {\bibfnamefont{D.~A.}\ \bibnamefont{Huse}},\ }%
  \bibfield{journal}{%
  \Doi{10.1103/PhysRevB.82.174411}{\bibinfo {journal} {Phys. Rev. B}}\ }%
  \textbf{\bibinfo {volume} {82}},\ \bibinfo {pages} {174411} (\bibinfo {year}
  {2010})%
  \bibAnnoteFile{NoStop}{pal_many-body_2010}%
\bibitem{imbrie}%
  \BibitemOpen
  \bibfield{author}{%
  \bibinfo {author} {\bibfnamefont{J.~Z.}\ \bibnamefont{Imbrie}},\ }%
  \bibfield{journal}{%
  \bibinfo {journal} {arXiv:1403.7837}}%
   (\bibinfo {year} {2014})%
  \bibAnnoteFile{NoStop}{imbrie}%
\bibitem{arcmp}%
  \BibitemOpen
  \bibfield{author}{%
  \bibinfo {author} {\bibfnamefont{R.}~\bibnamefont{Nandkishore}}\ and\
  \bibinfo {author} {\bibfnamefont{D.~A.}\ \bibnamefont{Huse}},\ }%
  \bibfield{journal}{%
  \bibinfo {journal} {arXiv:1404.0686}}%
   (\bibinfo {year} {2014})%
  \bibAnnoteFile{NoStop}{arcmp}%
\bibitem{huse_localization_2013}%
  \BibitemOpen
  \bibfield{author}{%
  \bibinfo {author} {\bibfnamefont{D.~A.}\ \bibnamefont{Huse}}, \bibinfo
  {author} {\bibfnamefont{R.}~\bibnamefont{Nandkishore}}, \bibinfo {author}
  {\bibfnamefont{V.}~\bibnamefont{Oganesyan}}, \bibinfo {author}
  {\bibfnamefont{A.}~\bibnamefont{Pal}},\ and\ \bibinfo {author}
  {\bibfnamefont{S.~L.}\ \bibnamefont{Sondhi}},\ }%
  \bibfield{journal}{%
  \Doi{10.1103/PhysRevB.88.014206}{\bibinfo {journal} {Phys. Rev. B}}\ }%
  \textbf{\bibinfo {volume} {88}},\ \bibinfo {pages} {014206} (\bibinfo {year}
  {2013})%
  \bibAnnoteFile{NoStop}{huse_localization_2013}%
\bibitem{pekker_hilbert-glass_2013}%
  \BibitemOpen
  \bibfield{author}{%
  \bibinfo {author} {\bibfnamefont{D.}~\bibnamefont{Pekker}}, \bibinfo {author}
  {\bibfnamefont{G.}~\bibnamefont{Refael}}, \bibinfo {author}
  {\bibfnamefont{E.}~\bibnamefont{Altman}},\ and\ \bibinfo {author}
  {\bibfnamefont{E.}~\bibnamefont{Demler}},\ }%
  \bibfield{journal}{%
  \bibinfo {journal} {arXiv:1307.3253}}%
   (\bibinfo {year} {2013})%
  \bibAnnoteFile{NoStop}{pekker_hilbert-glass_2013}%
\bibitem{bahri_localization_2013}%
  \BibitemOpen
  \bibfield{author}{%
  \bibinfo {author} {\bibfnamefont{Y.}~\bibnamefont{Bahri}}, \bibinfo {author}
  {\bibfnamefont{R.}~\bibnamefont{Vosk}}, \bibinfo {author}
  {\bibfnamefont{E.}~\bibnamefont{Altman}},\ and\ \bibinfo {author}
  {\bibfnamefont{A.}~\bibnamefont{Vishwanath}},\ }%
  \bibfield{journal}{%
  \bibinfo {journal} {arXiv:1307.4092}}%
   (\bibinfo {year} {2013})%
  \bibAnnoteFile{NoStop}{bahri_localization_2013}%
\bibitem{qhmbl}%
  \BibitemOpen
  \bibfield{author}{%
  \bibinfo {author} {\bibfnamefont{R.}~\bibnamefont{Nandkishore}}\ and\
  \bibinfo {author} {\bibfnamefont{A.~C.}\ \bibnamefont{Potter}},\ }%
  \bibfield{journal}{%
  \Doi{10.1103/PhysRevB.90.195115}{\bibinfo {journal} {Phys. Rev. B}}\ }%
  \textbf{\bibinfo {volume} {90}},\ \bibinfo {pages} {195115} (\bibinfo {month}
  {Nov}\ \bibinfo {year} {2014}),\
  \url{http://link.aps.org/doi/10.1103/PhysRevB.90.195115}%
  \bibAnnoteFile{NoStop}{qhmbl}%
\bibitem{vosk_many-body_2013}%
  \BibitemOpen
  \bibfield{author}{%
  \bibinfo {author} {\bibfnamefont{R.}~\bibnamefont{Vosk}}\ and\ \bibinfo
  {author} {\bibfnamefont{E.}~\bibnamefont{Altman}},\ }%
  \bibfield{journal}{%
  \Doi{10.1103/PhysRevLett.110.067204}{\bibinfo {journal} {Phys. Rev. Lett.}}\
  }%
  \textbf{\bibinfo {volume} {110}},\ \bibinfo {pages} {067204} (\bibinfo {year}
  {2013})%
  \bibAnnoteFile{NoStop}{vosk_many-body_2013}%
\bibitem{serbyn_universal_2013}%
  \BibitemOpen
  \bibfield{author}{%
  \bibinfo {author} {\bibfnamefont{M.}~\bibnamefont{Serbyn}}, \bibinfo {author}
  {\bibfnamefont{Z.}~\bibnamefont{Papi{\'c}}},\ and\ \bibinfo {author}
  {\bibfnamefont{D.~A.}\ \bibnamefont{Abanin}},\ }%
  \bibfield{journal}{%
  \Doi{10.1103/PhysRevLett.110.260601}{\bibinfo {journal} {Phys. Rev. Lett.}}\
  }%
  \textbf{\bibinfo {volume} {110}},\ \bibinfo {pages} {260601} (\bibinfo {year}
  {2013})%
  \bibAnnoteFile{NoStop}{serbyn_universal_2013}%
\bibitem{serbyn_local_2013}%
  \BibitemOpen
  \bibfield{author}{%
  \bibinfo {author} {\bibfnamefont{M.}~\bibnamefont{Serbyn}}, \bibinfo {author}
  {\bibfnamefont{Z.}~\bibnamefont{Papi{\'c}}},\ and\ \bibinfo {author}
  {\bibfnamefont{D.~A.}\ \bibnamefont{Abanin}},\ }%
  \bibfield{journal}{%
  \Doi{10.1103/PhysRevLett.111.127201}{\bibinfo {journal} {Phys. Rev. Lett.}}\
  }%
  \textbf{\bibinfo {volume} {111}},\ \bibinfo {pages} {127201} (\bibinfo {year}
  {2013})%
  \bibAnnoteFile{NoStop}{serbyn_local_2013}%
\bibitem{huse_phenomenology_2013}%
  \BibitemOpen
  \bibfield{author}{%
  \bibinfo {author} {\bibfnamefont{D.~A.}\ \bibnamefont{Huse}}\ and\ \bibinfo
  {author} {\bibfnamefont{V.}~\bibnamefont{Oganesyan}},\ }%
  \bibfield{journal}{%
  \bibinfo {journal} {{arXiv:1305.4915}}}%
   (\bibinfo {year} {2013})%
  \bibAnnoteFile{NoStop}{huse_phenomenology_2013}%
\bibitem{bauer_area_2013}%
  \BibitemOpen
  \bibfield{author}{%
  \bibinfo {author} {\bibfnamefont{B.}~\bibnamefont{Bauer}}\ and\ \bibinfo
  {author} {\bibfnamefont{C.}~\bibnamefont{Nayak}},\ }%
  \bibfield{journal}{%
  \bibinfo {journal} {arXiv:1306.5753}}%
   (\bibinfo {year} {2013})%
  \bibAnnoteFile{NoStop}{bauer_area_2013}%
\bibitem{huse_phenomenology_2014}%
  \BibitemOpen
  \bibfield{author}{%
  \bibinfo {author} {\bibfnamefont{D.~A.}\ \bibnamefont{Huse}}, \bibinfo
  {author} {\bibfnamefont{R.}~\bibnamefont{Nandkishore}},\ and\ \bibinfo
  {author} {\bibfnamefont{V.}~\bibnamefont{Oganesyan}},\ }%
  \bibfield{journal}{%
  \bibinfo {journal} {in preparation}}%
   (\bibinfo {year} {2014})%
  \bibAnnoteFile{NoStop}{huse_phenomenology_2014}%
\bibitem{kjall}%
  \BibitemOpen
  \bibfield{author}{%
  \bibinfo {author} {\bibfnamefont{J.~A.}\ \bibnamefont{Kj\"{a}ll}}, \bibinfo
  {author} {\bibfnamefont{J.~H.}\ \bibnamefont{Bardarson}},\ and\ \bibinfo
  {author} {\bibfnamefont{F.}~\bibnamefont{Pollmann}},\ }%
  \bibfield{journal}{%
  \bibinfo {journal} {arXiv:1403.1568}}%
   (\bibinfo {year} {2014})%
  \bibAnnoteFile{NoStop}{kjall}%
\bibitem{barlev1}%
  \BibitemOpen
  \bibfield{author}{%
  \bibinfo {author} {\bibfnamefont{Y.}~\bibnamefont{Bar~Lev}}\ and\ \bibinfo
  {author} {\bibfnamefont{D.~R.}\ \bibnamefont{Reichman}},\ }%
  \bibfield{journal}{%
  \Doi{10.1103/PhysRevB.89.220201}{\bibinfo {journal} {Phys. Rev. B}}\ }%
  \textbf{\bibinfo {volume} {89}},\ \bibinfo {pages} {220201} (\bibinfo {month}
  {Jun}\ \bibinfo {year} {2014}),\
  \url{http://link.aps.org/doi/10.1103/PhysRevB.89.220201}%
  \bibAnnoteFile{NoStop}{barlev1}%
\bibitem{barlev2}%
  \BibitemOpen
  \bibfield{author}{%
  \bibinfo {author} {\bibfnamefont{Y.}~\bibnamefont{Bar~Lev}}, \bibinfo
  {author} {\bibfnamefont{G.}~\bibnamefont{Cohen}},\ and\ \bibinfo {author}
  {\bibfnamefont{D.~R.}\ \bibnamefont{Reichman}},\ }%
  \bibfield{journal}{%
  \bibinfo {journal} {arXiv:1407.7535}}%
   (\bibinfo {year} {2014})%
  \bibAnnoteFile{NoStop}{barlev2}%
\bibitem{kartiek}%
  \BibitemOpen
  \bibfield{author}{%
  \bibinfo {author} {\bibfnamefont{K.}~\bibnamefont{Agarwal}}, \bibinfo
  {author} {\bibfnamefont{S.}~\bibnamefont{Gopalakrishnan}}, \bibinfo {author}
  {\bibfnamefont{M.}~\bibnamefont{Knap}}, \bibinfo {author}
  {\bibfnamefont{M.}~\bibnamefont{Mueller}},\ and\ \bibinfo {author}
  {\bibfnamefont{E.}~\bibnamefont{Demler}},\ }%
  \bibfield{journal}{%
  \bibinfo {journal} {arXiv:1408.3413}}%
   (\bibinfo {year} {2014})%
  \bibAnnoteFile{NoStop}{kartiek}%
\bibitem{laflorencie}%
  \BibitemOpen
  \bibfield{author}{%
  \bibinfo {author} {\bibfnamefont{D.~J.}\ \bibnamefont{Luitz}}, \bibinfo
  {author} {\bibfnamefont{N.}~\bibnamefont{Laflorencie}},\ and\ \bibinfo
  {author} {\bibfnamefont{F.}~\bibnamefont{Alet}},\ }%
  \bibfield{journal}{%
  \bibinfo {journal} {arXiv:1411.0660}}%
   (\bibinfo {year} {2014})%
  \bibAnnoteFile{NoStop}{laflorencie}%
\bibitem{kotliar_RMP}%
  \BibitemOpen
  \bibfield{author}{%
  \bibinfo {author} {\bibfnamefont{A.}~\bibnamefont{Georges}}, \bibinfo
  {author} {\bibfnamefont{G.}~\bibnamefont{Kotliar}}, \bibinfo {author}
  {\bibfnamefont{W.}~\bibnamefont{Krauth}},\ and\ \bibinfo {author}
  {\bibfnamefont{M.~J.}\ \bibnamefont{Rozenberg}},\ }%
  \bibfield{journal}{%
  \Doi{10.1103/RevModPhys.68.13}{\bibinfo {journal} {Rev. Mod. Phys.}}\ }%
  \textbf{\bibinfo {volume} {68}},\ \bibinfo {pages} {13} (\bibinfo {month}
  {Jan}\ \bibinfo {year} {1996}),\
  \url{http://link.aps.org/doi/10.1103/RevModPhys.68.13}%
  \bibAnnoteFile{NoStop}{kotliar_RMP}%
\bibitem{dobr}%
  \BibitemOpen
  \bibfield{author}{%
  \bibinfo {author} {\bibfnamefont{V.}~\bibnamefont{Dobrosavljevic}}, \bibinfo
  {author} {\bibfnamefont{A.~A.}\ \bibnamefont{Pastor}},\ and\ \bibinfo
  {author} {\bibfnamefont{B.}~\bibnamefont{Nikolic}},\ }%
  \bibfield{journal}{%
  \bibinfo {journal} {Europhys. Lett.}\ }%
  \textbf{\bibinfo {volume} {62}},\ \bibinfo {pages} {76} (\bibinfo {year}
  {2003})%
  \bibAnnoteFile{NoStop}{dobr}%
\bibitem{dobr2}%
  \BibitemOpen
  \bibfield{author}{%
  \bibinfo {author} {\bibfnamefont{M.~C.~O.}\ \bibnamefont{Aguiar}}, \bibinfo
  {author} {\bibfnamefont{V.}~\bibnamefont{Dobrosavljevi\ifmmode~\acute{c}\else
  \'{c}\fi{}}}, \bibinfo {author} {\bibfnamefont{E.}~\bibnamefont{Abrahams}},\
  and\ \bibinfo {author} {\bibfnamefont{G.}~\bibnamefont{Kotliar}},\ }%
  \bibfield{journal}{%
  \Doi{10.1103/PhysRevLett.102.156402}{\bibinfo {journal} {Phys. Rev. Lett.}}\
  }%
  \textbf{\bibinfo {volume} {102}},\ \bibinfo {pages} {156402} (\bibinfo
  {month} {Apr}\ \bibinfo {year} {2009}),\
  \url{http://link.aps.org/doi/10.1103/PhysRevLett.102.156402}%
  \bibAnnoteFile{NoStop}{dobr2}%
\bibitem{ngh}%
  \BibitemOpen
  \bibfield{author}{%
  \bibinfo {author} {\bibfnamefont{R.}~\bibnamefont{Nandkishore}}, \bibinfo
  {author} {\bibfnamefont{S.}~\bibnamefont{Gopalakrishnan}},\ and\ \bibinfo
  {author} {\bibfnamefont{D.~A.}\ \bibnamefont{Huse}},\ }%
  \bibinfo {journal} {arXiv:1402.5971}%
  \bibAnnoteFile{NoStop}{ngh}%
\bibitem{levitov1990}%
  \BibitemOpen
\bibfield{journal}{%
    }%
  \bibfield{author}{%
  \bibinfo {author} {\bibfnamefont{L.~S.}\ \bibnamefont{Levitov}},\ }%
  \bibfield{journal}{%
  \bibinfo {journal} {Phys. Rev. Lett.}\ }%
  \textbf{\bibinfo {volume} {64}},\ \bibinfo {pages} {547} (\bibinfo {year}
  {1990})%
  \bibAnnoteFile{NoStop}{levitov1990}%
\bibitem{burin98}%
  \BibitemOpen
  \bibfield{author}{%
  \bibinfo {author} {\bibfnamefont{A.~L.}\ \bibnamefont{Burin}}, \bibinfo
  {author} {\bibfnamefont{Y.}~\bibnamefont{Kagan}}, \bibinfo {author}
  {\bibfnamefont{L.~A.}\ \bibnamefont{Maksimov}},\ and\ \bibinfo {author}
  {\bibfnamefont{I.~Y.}\ \bibnamefont{Polishchuk}},\ }%
  \bibfield{journal}{%
  \Doi{10.1103/PhysRevLett.80.2945}{\bibinfo {journal} {Phys. Rev. Lett.}}\ }%
  \textbf{\bibinfo {volume} {80}},\ \bibinfo {pages} {2945} (\bibinfo {month}
  {Mar}\ \bibinfo {year} {1998}),\
  \url{http://link.aps.org/doi/10.1103/PhysRevLett.80.2945}%
  \bibAnnoteFile{NoStop}{burin98}%
\bibitem{yao13}%
  \BibitemOpen
  \bibfield{author}{%
  \bibinfo {author} {\bibfnamefont{N.~Y.}\ \bibnamefont{{Yao}}}, \bibinfo
  {author} {\bibfnamefont{C.~R.}\ \bibnamefont{{Laumann}}}, \bibinfo {author}
  {\bibfnamefont{S.}~\bibnamefont{{Gopalakrishnan}}}, \bibinfo {author}
  {\bibfnamefont{M.}~\bibnamefont{{Knap}}}, \bibinfo {author}
  {\bibfnamefont{M.}~\bibnamefont{{Mueller}}}, \bibinfo {author}
  {\bibfnamefont{E.~A.}\ \bibnamefont{{Demler}}},\ and\ \bibinfo {author}
  {\bibfnamefont{M.~D.}\ \bibnamefont{{Lukin}}},\ }%
  \bibfield{journal}{%
  \bibinfo {journal} {ArXiv e-prints}}%
   (\bibinfo {month} {Nov.}\ \bibinfo {year} {2013}),\
  \Eprint{http://arxiv.org/abs/1311.7151}{arXiv:1311.7151
  [cond-mat.stat-mech]}%
  \bibAnnoteFile{NoStop}{yao13}%
\bibitem{agkl}%
  \BibitemOpen
  \bibfield{author}{%
  \bibinfo {author} {\bibfnamefont{B.~L.}\ \bibnamefont{Altshuler}}, \bibinfo
  {author} {\bibfnamefont{Y.}~\bibnamefont{Gefen}}, \bibinfo {author}
  {\bibfnamefont{A.}~\bibnamefont{Kamenev}},\ and\ \bibinfo {author}
  {\bibfnamefont{L.~S.}\ \bibnamefont{Levitov}},\ }%
  \bibfield{journal}{%
  \Doi{10.1103/PhysRevLett.78.2803}{\bibinfo {journal} {Phys. Rev. Lett.}}\ }%
  \textbf{\bibinfo {volume} {78}},\ \bibinfo {pages} {2803} (\bibinfo {year}
  {1997})%
  \bibAnnoteFile{NoStop}{agkl}%
\bibitem{laumann2014}%
  \BibitemOpen
  \bibfield{author}{%
  \bibinfo {author} {\bibfnamefont{C.}~\bibnamefont{Laumann}}, \bibinfo
  {author} {\bibfnamefont{A.}~\bibnamefont{Pal}},\ and\ \bibinfo {author}
  {\bibfnamefont{A.}~\bibnamefont{Scardicchio}},\ }%
  \bibfield{journal}{%
  \bibinfo {journal} {arXiv:1404.2276}}%
   (\bibinfo {year} {2014})%
  \bibAnnoteFile{NoStop}{laumann2014}%
\bibitem{Note1}%
  \BibitemOpen
  \bibinfo {note} {We note that there are some similarities between Section IV
  and Ref.\protect \rev@citealpnum {Fiegelman}. However, \protect
  \rev@citealpnum {Fiegelman} is concerned with the localization transition for
  single excitations above the ground state, whereas we are concerned with the
  very different problem of the localization transition for generic high energy
  states. Moreover, unlike \protect \rev@citealpnum {Fiegelman} we work on a
  regular lattice (not a Bethe lattice), and incorporate Hartree terms; as we
  discuss, this changes the nature of metallic-state solutions.}%
  \bibAnnoteFile{Stop}{Note1}%
\bibitem{mottVRH}%
  \BibitemOpen
  \bibfield{author}{%
  \bibinfo {author} {\bibfnamefont{N.~F.}\ \bibnamefont{Mott}},\ }%
  \bibfield{journal}{%
  \bibinfo {journal} {J. Non-Crystalline Solids}\ }%
  \textbf{\bibinfo {volume} {1}},\ \bibinfo {pages} {1} (\bibinfo {year}
  {1968})%
  \bibAnnoteFile{NoStop}{mottVRH}%
\bibitem{mblDeer}%
  \BibitemOpen
  \bibfield{author}{%
  \bibinfo {author} {\bibfnamefont{M.}~\bibnamefont{Serbyn}} \emph{et~al.},\ }%
  \bibinfo {journal} {arXiv:1403.0693}%
  \bibAnnoteFile{NoStop}{mblDeer}%
\bibitem{galperin}%
  \BibitemOpen
\bibfield{journal}{%
    }%
  \bibfield{author}{%
  \bibinfo {author} {\bibfnamefont{Y.}~\bibnamefont{Galperin}}, \bibinfo
  {author} {\bibfnamefont{B.}~\bibnamefont{Altshuler}}, \bibinfo {author}
  {\bibfnamefont{J.}~\bibnamefont{Bergli}},\ and\ \bibinfo {author}
  {\bibfnamefont{D.}~\bibnamefont{Shantsev}},\ }%
  \bibfield{journal}{%
  \Doi{10.1103/PhysRevLett.96.097009}{\bibinfo {journal} {Phys. Rev. Lett.}}\
  }%
  \textbf{\bibinfo {volume} {96}},\ \bibinfo {pages} {097009} (\bibinfo {month}
  {Mar}\ \bibinfo {year} {2006}),\
  \url{http://link.aps.org/doi/10.1103/PhysRevLett.96.097009}%
  \bibAnnoteFile{NoStop}{galperin}%
\bibitem{khaetskii1}%
  \BibitemOpen
  \bibfield{author}{%
  \bibinfo {author} {\bibfnamefont{A.~V.}\ \bibnamefont{Khaetskii}}, \bibinfo
  {author} {\bibfnamefont{D.}~\bibnamefont{Loss}},\ and\ \bibinfo {author}
  {\bibfnamefont{L.}~\bibnamefont{Glazman}},\ }%
  \bibfield{journal}{%
  \Doi{10.1103/PhysRevLett.88.186802}{\bibinfo {journal} {Phys. Rev. Lett.}}\
  }%
  \textbf{\bibinfo {volume} {88}},\ \bibinfo {pages} {186802} (\bibinfo {month}
  {Apr}\ \bibinfo {year} {2002}),\
  \url{http://link.aps.org/doi/10.1103/PhysRevLett.88.186802}%
  \bibAnnoteFile{NoStop}{khaetskii1}%
\bibitem{khaetskii2}%
  \BibitemOpen
  \bibfield{author}{%
  \bibinfo {author} {\bibfnamefont{A.}~\bibnamefont{Khaetskii}}, \bibinfo
  {author} {\bibfnamefont{D.}~\bibnamefont{Loss}},\ and\ \bibinfo {author}
  {\bibfnamefont{L.}~\bibnamefont{Glazman}},\ }%
  \bibfield{journal}{%
  \Doi{10.1103/PhysRevB.67.195329}{\bibinfo {journal} {Phys. Rev. B}}\ }%
  \textbf{\bibinfo {volume} {67}},\ \bibinfo {pages} {195329} (\bibinfo {month}
  {May}\ \bibinfo {year} {2003}),\
  \url{http://link.aps.org/doi/10.1103/PhysRevB.67.195329}%
  \bibAnnoteFile{NoStop}{khaetskii2}%
\bibitem{witzel1}%
  \BibitemOpen
  \bibfield{author}{%
  \bibinfo {author} {\bibfnamefont{W.~M.}\ \bibnamefont{Witzel}}\ and\ \bibinfo
  {author} {\bibfnamefont{S.}~\bibnamefont{Das~Sarma}},\ }%
  \bibfield{journal}{%
  \Doi{10.1103/PhysRevB.74.035322}{\bibinfo {journal} {Phys. Rev. B}}\ }%
  \textbf{\bibinfo {volume} {74}},\ \bibinfo {pages} {035322} (\bibinfo {month}
  {Jul}\ \bibinfo {year} {2006}),\
  \url{http://link.aps.org/doi/10.1103/PhysRevB.74.035322}%
  \bibAnnoteFile{NoStop}{witzel1}%
\bibitem{witzel2}%
  \BibitemOpen
  \bibfield{author}{%
  \bibinfo {author} {\bibfnamefont{L.}~\bibnamefont{Cywi\ifmmode~\acute{n}\else
  \'{n}\fi{}ski}}, \bibinfo {author} {\bibfnamefont{W.~M.}\
  \bibnamefont{Witzel}},\ and\ \bibinfo {author}
  {\bibfnamefont{S.}~\bibnamefont{Das~Sarma}},\ }%
  \bibfield{journal}{%
  \Doi{10.1103/PhysRevLett.102.057601}{\bibinfo {journal} {Phys. Rev. Lett.}}\
  }%
  \textbf{\bibinfo {volume} {102}},\ \bibinfo {pages} {057601} (\bibinfo
  {month} {Feb}\ \bibinfo {year} {2009}),\
  \url{http://link.aps.org/doi/10.1103/PhysRevLett.102.057601}%
  \bibAnnoteFile{NoStop}{witzel2}%
\bibitem{bloembergen}%
  \BibitemOpen
  \bibfield{author}{%
  \bibinfo {author} {\bibfnamefont{N.}~\bibnamefont{Bloembergen}},\ }%
  \bibfield{journal}{%
  \bibinfo {journal} {Physica}\ }%
  \textbf{\bibinfo {volume} {15}},\ \bibinfo {pages} {386} (\bibinfo {year}
  {1949})%
  \bibAnnoteFile{NoStop}{bloembergen}%
\bibitem{szabo}%
  \BibitemOpen
  \bibfield{author}{%
  \bibinfo {author} {\bibfnamefont{A.}~\bibnamefont{Szabo}}, \bibinfo {author}
  {\bibfnamefont{T.}~\bibnamefont{Muramoto}},\ and\ \bibinfo {author}
  {\bibfnamefont{R.}~\bibnamefont{Kaarli}},\ }%
  \bibfield{journal}{%
  \Doi{10.1103/PhysRevB.42.7769}{\bibinfo {journal} {Phys. Rev. B}}\ }%
  \textbf{\bibinfo {volume} {42}},\ \bibinfo {pages} {7769} (\bibinfo {month}
  {Nov}\ \bibinfo {year} {1990}),\
  \url{http://link.aps.org/doi/10.1103/PhysRevB.42.7769}%
  \bibAnnoteFile{NoStop}{szabo}%
\bibitem{paola2009}%
  \BibitemOpen
  \bibfield{author}{%
  \bibinfo {author} {\bibfnamefont{P.}~\bibnamefont{Cappellaro}}, \bibinfo
  {author} {\bibfnamefont{L.}~\bibnamefont{Jiang}}, \bibinfo {author}
  {\bibfnamefont{J.~S.}\ \bibnamefont{Hodges}},\ and\ \bibinfo {author}
  {\bibfnamefont{M.~D.}\ \bibnamefont{Lukin}},\ }%
  \bibfield{journal}{%
  \Doi{10.1103/PhysRevLett.102.210502}{\bibinfo {journal} {Phys. Rev. Lett.}}\
  }%
  \textbf{\bibinfo {volume} {102}},\ \bibinfo {pages} {210502} (\bibinfo
  {month} {May}\ \bibinfo {year} {2009}),\
  \url{http://link.aps.org/doi/10.1103/PhysRevLett.102.210502}%
  \bibAnnoteFile{NoStop}{paola2009}%
\bibitem{ariel_phonon}%
  \BibitemOpen
  \bibfield{author}{%
  \bibinfo {author} {\bibfnamefont{A.}~\bibnamefont{Amir}}, \bibinfo {author}
  {\bibfnamefont{J.~J.}\ \bibnamefont{Krich}}, \bibinfo {author}
  {\bibfnamefont{V.}~\bibnamefont{Vitelli}}, \bibinfo {author}
  {\bibfnamefont{Y.}~\bibnamefont{Oreg}},\ and\ \bibinfo {author}
  {\bibfnamefont{Y.}~\bibnamefont{Imry}},\ }%
  \bibfield{journal}{%
  \Doi{10.1103/PhysRevX.3.021017}{\bibinfo {journal} {Phys. Rev. X}}\ }%
  \textbf{\bibinfo {volume} {3}},\ \bibinfo {pages} {021017} (\bibinfo {month}
  {Jun}\ \bibinfo {year} {2013}),\
  \url{http://link.aps.org/doi/10.1103/PhysRevX.3.021017}%
  \bibAnnoteFile{NoStop}{ariel_phonon}%
\bibitem{ariel-diffusion}%
  \BibitemOpen
  \bibfield{author}{%
  \bibinfo {author} {\bibfnamefont{A.}~\bibnamefont{Amir}}, \bibinfo {author}
  {\bibfnamefont{Y.}~\bibnamefont{Lahini}},\ and\ \bibinfo {author}
  {\bibfnamefont{H.~B.}\ \bibnamefont{Perets}},\ }%
  \bibfield{journal}{%
  \Doi{10.1103/PhysRevE.79.050105}{\bibinfo {journal} {Phys. Rev. E}}\ }%
  \textbf{\bibinfo {volume} {79}},\ \bibinfo {pages} {050105} (\bibinfo {month}
  {May}\ \bibinfo {year} {2009}),\
  \url{http://link.aps.org/doi/10.1103/PhysRevE.79.050105}%
  \bibAnnoteFile{NoStop}{ariel-diffusion}%
\bibitem{Note2}%
  \BibitemOpen
  \bibinfo {note} {A. Amir, in preparation}%
  \bibAnnoteFile{NoStop}{Note2}%
\bibitem{Note3}%
  \BibitemOpen
  \bibinfo {note} {In a homogeneous system, the MBL transition occurs at $g
  \protect \alt 1$; however, as we are considering a system of coupled grains,
  we may consistently assume that $g \gg 1$.}%
  \bibAnnoteFile{Stop}{Note3}%
\bibitem{aleiner-review}%
  \BibitemOpen
  \bibfield{author}{%
  \bibinfo {author} {\bibfnamefont{I.~L.}\ \bibnamefont{Aleiner}}, \bibinfo
  {author} {\bibfnamefont{P.~W.}\ \bibnamefont{Brouwer}},\ and\ \bibinfo
  {author} {\bibfnamefont{L.~I.}\ \bibnamefont{Glazman}},\ }%
  \bibfield{journal}{%
  \bibinfo {journal} {Phys. Rep.}\ }%
  \textbf{\bibinfo {volume} {358}},\ \bibinfo {pages} {309} (\bibinfo {year}
  {2002})%
  \bibAnnoteFile{NoStop}{aleiner-review}%
\bibitem{Note4}%
  \BibitemOpen
  \bibinfo {note} {Note that, although the \protect \emph {single-particle}
  properties are described by random-matrix theory, in general the \protect
  \emph {many-body} level statistics are not.}%
  \bibAnnoteFile{Stop}{Note4}%
\bibitem{abouchacra}%
  \BibitemOpen
  \bibfield{author}{%
  \bibinfo {author} {\bibfnamefont{R.}~\bibnamefont{Abou-Chacra}}, \bibinfo
  {author} {\bibfnamefont{P.~W.}\ \bibnamefont{Anderson}},\ and\ \bibinfo
  {author} {\bibfnamefont{D.~J.}\ \bibnamefont{Thouless}},\ }%
  \bibfield{journal}{%
  \bibinfo {journal} {J. Phys. C}\ }%
  \textbf{\bibinfo {volume} {6}},\ \bibinfo {pages} {1734} (\bibinfo {year}
  {1973})%
  \bibAnnoteFile{NoStop}{abouchacra}%
\bibitem{chalker}%
  \BibitemOpen
  \bibfield{author}{%
  \bibinfo {author} {\bibfnamefont{J.~T.}\ \bibnamefont{Chalker}}\ and\
  \bibinfo {author} {\bibfnamefont{S.}~\bibnamefont{Siak}},\ }%
  \bibfield{journal}{%
  \bibinfo {journal} {J. Phys.: Condensed Matter}\ }%
  \textbf{\bibinfo {volume} {2}},\ \bibinfo {pages} {2671} (\bibinfo {year}
  {1990})%
  \bibAnnoteFile{NoStop}{chalker}%
\bibitem{beenakker}%
  \BibitemOpen
  \bibfield{author}{%
  \bibinfo {author} {\bibfnamefont{X.}~\bibnamefont{Leyronas}}, \bibinfo
  {author} {\bibfnamefont{P.~G.}\ \bibnamefont{Silvestrov}},\ and\ \bibinfo
  {author} {\bibfnamefont{C.~W.~J.}\ \bibnamefont{Beenakker}},\ }%
  \bibfield{journal}{%
  \Doi{10.1103/PhysRevLett.84.3414}{\bibinfo {journal} {Phys. Rev. Lett.}}\ }%
  \textbf{\bibinfo {volume} {84}},\ \bibinfo {pages} {3414} (\bibinfo {month}
  {Apr}\ \bibinfo {year} {2000}),\
  \url{http://link.aps.org/doi/10.1103/PhysRevLett.84.3414}%
  \bibAnnoteFile{NoStop}{beenakker}%
\bibitem{kamenevQD}%
  \BibitemOpen
  \bibfield{author}{%
  \bibinfo {author} {\bibfnamefont{A.~M.~F.}\ \bibnamefont{Rivas}}, \bibinfo
  {author} {\bibfnamefont{E.~R.}\ \bibnamefont{Mucciolo}},\ and\ \bibinfo
  {author} {\bibfnamefont{A.}~\bibnamefont{Kamenev}},\ }%
  \bibfield{journal}{%
  \Doi{10.1103/PhysRevB.65.155309}{\bibinfo {journal} {Phys. Rev. B}}\ }%
  \textbf{\bibinfo {volume} {65}},\ \bibinfo {pages} {155309} (\bibinfo {month}
  {Mar}\ \bibinfo {year} {2002}),\
  \url{http://link.aps.org/doi/10.1103/PhysRevB.65.155309}%
  \bibAnnoteFile{NoStop}{kamenevQD}%
\bibitem{Note5}%
  \BibitemOpen
  \bibinfo {note} {The l-bit model of the FMBL state is unsuitable for our
  present purposes, because it has a decay length for the interactions which
  diverges near the localization transition, and we want a model where the
  interactions are short range.}%
  \bibAnnoteFile{Stop}{Note5}%
\bibitem{Fiegelman}%
  \BibitemOpen
  \bibfield{author}{%
  \bibinfo {author} {\bibfnamefont{M.~V.}\ \bibnamefont{Feigel'man}}, \bibinfo
  {author} {\bibfnamefont{L.~B.}\ \bibnamefont{Ioffe}},\ and\ \bibinfo {author}
  {\bibfnamefont{M.}~\bibnamefont{M\'ezard}},\ }%
  \bibfield{journal}{%
  \Doi{10.1103/PhysRevB.82.184534}{\bibinfo {journal} {Phys. Rev. B}}\ }%
  \textbf{\bibinfo {volume} {82}},\ \bibinfo {pages} {184534} (\bibinfo {month}
  {Nov}\ \bibinfo {year} {2010}),\
  \url{http://link.aps.org/doi/10.1103/PhysRevB.82.184534}%
  \bibAnnoteFile{NoStop}{Fiegelman}%
\bibitem{Note6}%
  \BibitemOpen
  \bibinfo {note} {Taking into account the definition of $\Xi = 1/\protect
  \qopname \relax o{ln}(g \delta _m / \lambda \Delta )$, we see that the
  condition for validity of the above leading order in $\gamma $ approximation
  is $ \protect \frac {g \gamma }{\lambda \Delta } \ll 1$, i.e., the inter-dot
  coupling $\gamma $ must be weak enough}%
  \bibAnnoteFile{NoStop}{Note6}%
\bibitem{randomgraph}%
  \BibitemOpen
  \bibfield{author}{%
  \bibinfo {author} {\bibfnamefont{A.}~\bibnamefont{{De Luca}}}, \bibinfo
  {author} {\bibfnamefont{B.~L.}\ \bibnamefont{{Altshuler}}}, \bibinfo {author}
  {\bibfnamefont{V.~E.}\ \bibnamefont{{Kravtsov}}},\ and\ \bibinfo {author}
  {\bibfnamefont{A.}~\bibnamefont{{Scardicchio}}},\ }%
  \bibfield{journal}{%
  \bibinfo {journal} {arxiv:1403.7817}}%
   (\bibinfo {year} {2014})%
  \bibAnnoteFile{NoStop}{randomgraph}%
\bibitem{grover}%
  \BibitemOpen
  \bibfield{author}{%
  \bibinfo {author} {\bibfnamefont{T.}~\bibnamefont{Grover}},\ }%
  \bibfield{journal}{%
  \bibinfo {journal} {arXiv:1405.1471}}%
   (\bibinfo {year} {2014})%
  \bibAnnoteFile{NoStop}{grover}%
\end{thebibliography}

%

\end{document}